\documentclass[superscriptaddress, twocolumn, prx, longbibliography, nofootinbib]{revtex4-1}
\usepackage{amsmath, amssymb, color, stmaryrd, wasysym, esint}
\makeatletter
\newsavebox{\@brx}
\newcommand{\llangle}[1][]{\savebox{\@brx}{\(\m@th{#1\langle}\)}%
  \mathopen{\copy\@brx\kern-0.5\wd\@brx\usebox{\@brx}}}
\newcommand{\rrangle}[1][]{\savebox{\@brx}{\(\m@th{#1\rangle}\)}%
  \mathclose{\copy\@brx\kern-0.5\wd\@brx\usebox{\@brx}}}
\makeatother

\makeatletter
\newsavebox{\@brxx}
\newcommand{\lllangle}[1][]{\savebox{\@brxx}{\(\m@th{#1\langle}\)}%
  \mathopen{\copy\@brxx\kern-0.5\wd\@brxx\usebox{\@brxx}\kern-0.5\wd\@brxx\usebox{\@brxx}}}
\newcommand{\rrrangle}[1][]{\savebox{\@brxx}{\(\m@th{#1\rangle}\)}%
  \mathclose{\copy\@brxx\kern-0.5\wd\@brxx\usebox{\@brxx}\kern-0.5\wd\@brxx\usebox{\@brxx}}}
\makeatother

\usepackage{graphicx}% Include figure files
\usepackage{dcolumn}% Align table columns on decimal point
\usepackage{bm}% bold math
\usepackage{xcolor}
\usepackage{xspace}
\usepackage[makeroom]{cancel}
\usepackage{float}
\usepackage{siunitx}
\usepackage{lineno}

\newcommand{\srangle}{\ensuremath{\rangle_\text{s}}}

\newcommand{\ssum}[3]{\ensuremath{\sum_{#1,#2\neq #1}}}

\definecolor{linkcolor}{rgb}{0,0,0.6} 
\usepackage[pdftex,colorlinks=true,
	pdfstartview = FitV,
	linkcolor    = linkcolor,
	citecolor    = linkcolor,
	urlcolor     = linkcolor,
	hyperindex   = true,
	hyperfigures = false]{hyperref}

\begin{document}

% ===========================================================================================

\title{Active matter under control: Insights from response theory}% Force line breaks with \\

\author{Luke K. Davis}
\email{luke.davis@ucl.ac.uk}
\affiliation{%
Department of Physics and Materials Science, University of Luxembourg, L-1511 Luxembourg
}%
\affiliation{%
Department of Mathematics, University College London, 25 Gordon Street, London, England
}%
\author{Karel Proesmans}
\email{karel.proesmans@nbi.ku.dk}
\affiliation{%
Department of Physics and Materials Science, University of Luxembourg, L-1511 Luxembourg
}%
\affiliation{%
Niels Bohr International Academy, Niels Bohr Institute, University of Copenhagen, Blegdamsvej 17, 2100 Copenhagen, Denmark
}%
\author{\'Etienne Fodor}%
\email{etienne.fodor@uni.lu}
\affiliation{%
Department of Physics and Materials Science, University of Luxembourg, L-1511 Luxembourg
}%

\begin{abstract}
Active constituents burn fuel to sustain individual motion, giving rise to collective effects that are not seen in systems at thermal equilibrium, such as phase separation with purely repulsive interactions. There is a great potential in harnessing the striking phenomenology of active matter to build novel controllable and responsive materials that surpass passive ones. Yet, we currently lack a systematic roadmap to predict the protocols driving active systems between different states in a way that is thermodynamically optimal. Equilibrium thermodynamics is an inadequate foundation to this end, due to the dissipation rate arising from the constant fuel consumption in active matter. Here, we derive and implement a versatile framework for the thermodynamic control of active matter. Combining recent developments in stochastic thermodynamics and response theory, our approach shows how to find the optimal control for either continuous- or discrete-state active systems operating out-of-equilibrium. Our results open the door to designing novel active materials which are not only built to stabilize specific nonequilibrium collective states, but are also optimized to switch between different states at minimum dissipation.
\end{abstract}

\maketitle
%\linenumbers

% ===========================================================================================

\section{Introduction}

Active matter is the class of nonequilibrium systems where individual constituents constantly consume energy to sustain directed motion~\cite{Marchetti2013, Gompper2020, fodor2021irreversibility, Wijland2022}. It encompasses both living~\cite{Prost2015,Needleman2017} and synthetic systems~\cite{Paxton2004, Howse2007, Palacci2013, Zottl2016, Vutukuri2020}, which can exhibit emergent phenomena not found at thermal equilibrium. Such emergent --collective-- properties are plentiful, ranging from the flocking of animals~\cite{Toner2005, Ballerini2008, Katz2011}, active turbulence~\cite{Alert2022}, swarming bacteria~\cite{Darnton2010, Beer2019}, to motility-induced phase separation (MIPS) that occurs in the absence of attractive particle interactions~\cite{Cates2015}.

Whilst the collective effects of active matter have been studied extensively, how to efficiently control active matter has only recently started to receive growing attention. Indeed, experiments have now  demonstrated the ability to manipulate active nematics with a magnetic field~\cite{Guillamat2016}, trigger spatial phase separation in bacteria~\cite{Fragkopoulos2020}, guide rotational patterns in magnetic rotors~\cite{Matsunaga2019}, and drive phase transitions in living cells~\cite{Shin2017}. Moreover, the interplay of external control and activity has shown to induce interesting nonlinear behaviours, such as negative mobility, where optimal protocols could result in novel collective functions \cite{Rizkallah2023}. Thus, the control of active matter opens unprecedented perspectives on developing novel nonequilibrium materials which selectively change collective states in response to perturbations. To this end, a first step is to establish generic principles for the reliable and optimal control of active systems.

Previous studies have explored the optimal control of active systems through the lens of minimal models. Such works have considered wet one-body control~\cite{Nasouri2021, DaddiMoussaIder2021, DaddiMoussaIder2021a}, one-body navigation strategies~\cite{Schneider2019, Liebchen2019, Piro2021, DaddiMoussaIder2021, Zanovello2021}, and more complex many-body scenarios and field theories~\cite{Norton2020, Chennakesavalu2021, Cavagna2021, Falk2021, Shankar2022}. In these works, optimality typically involves penalizing protocols that steer away from the target state and/or rewards those finishing in the least time, without any constraint on how much energy is dissipated. Therefore, how to optimize control under thermodynamic constraints, such as minimizing the dissipated heat, is still largely unexplored for active systems.

\begin{figure}[b!]
\includegraphics[width=0.35\textwidth]{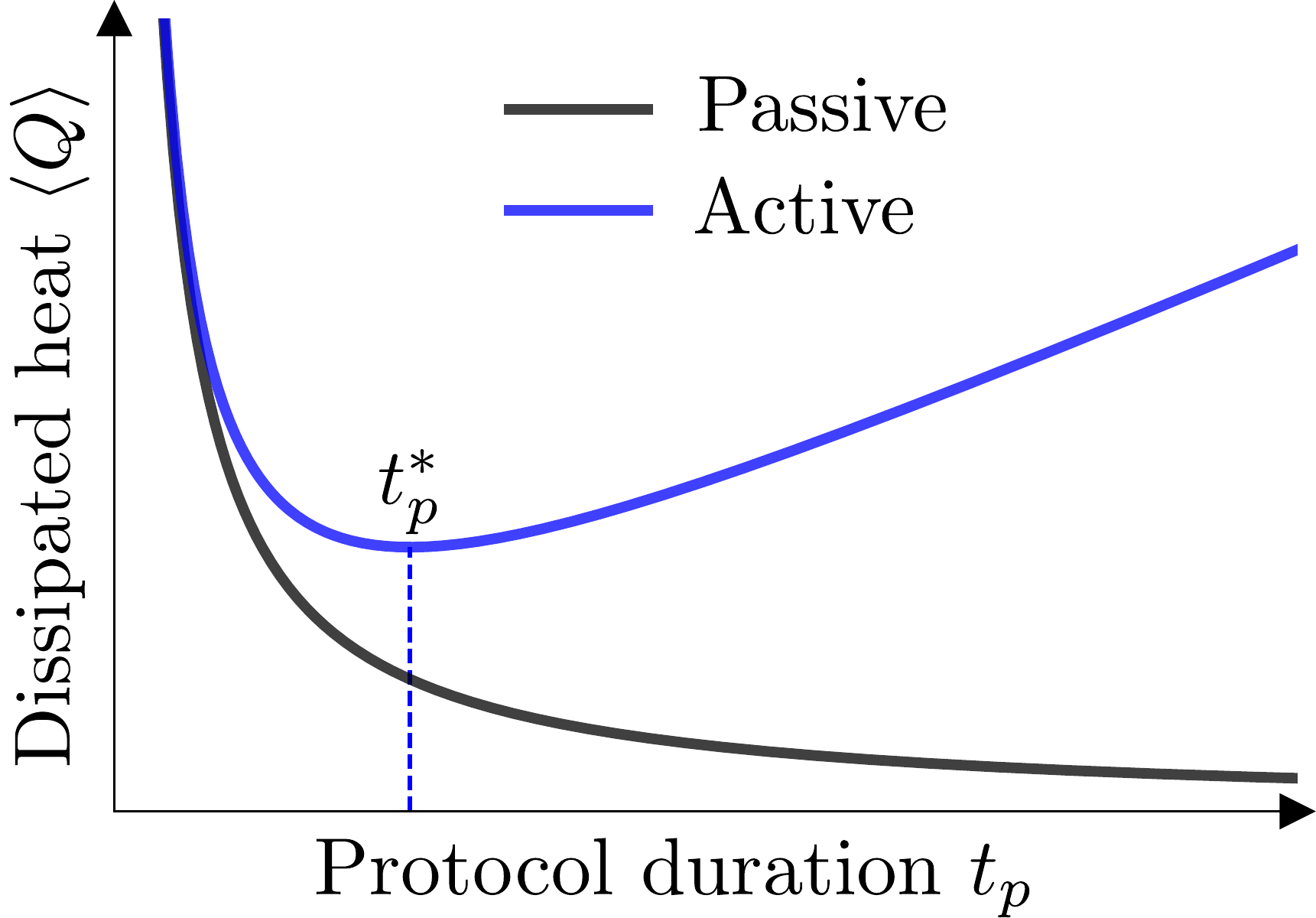}
\caption{\label{fig:Sketch} Illustration of the heat dissipated for a control protocol with external perturbation in finite time. In passive systems, the protocol achieving the least dissipation is always the slowest one, as expected from standard thermodynamics. In active systems, the optimal protocol has instead a finite duration $t_p^*$ which achieves the best trade-off between the dissipation stemming from the external perturbation, and that coming from internal activity.}
\end{figure}

For passive systems, recent advances in stochastic thermodynamics have led to a versatile framework for control at minimum dissipation~\cite{Crooks2007, Sivak2012, Rotskoff2015, Rotskoff2017, Tafoya2019, Blaber2020, Deffner2020}. In this context, the operator controls the parameters of a given potential energy. The corresponding dissipation decreases with protocol time [Fig.~\ref{fig:Sketch}], so that the slowest protocol is always optimal, as expected from the thermodynamics of passive systems. For passive systems, the control framework amounts to geometric optimization~\cite{Crooks2007,Deffner2020}: optimal trajectories are geodesics on curved manifolds of a thermodynamic metric. The success of this geometric approach \cite{Rotskoff2015,Rotskoff2017,Blaber2020,Large2019,Frim2022} fosters the hope that the underlying framework could also inform control beyond inherently passive systems. How such a framework can be adapted to active matter remains to be explored.

Underpinning the geometric approach for optimal control is linear response theory. It allows one to predict the effect of a weak perturbation in terms of correlations in the unperturbed dynamics. For passive systems, the relation between response and correlation functions follows the celebrated fluctuation-dissipation theorem~\cite{Kubo1966}, which relies on the time-reversal symmetry (TRS) of the dynamics. Interestingly, such a relation can be straightforwardly extended when TRS is broken~\cite{Marconi2008, Baiesi2009}. Indeed, it has already been used to study the response of various active systems~\cite{DalCengio2019, Maes2020, Martin2021, DalCengio2021, Caprini2021}. These recent developments establish a clear roadmap to extend the framework of thermodynamic control from passive to active systems.

In active systems, the energy consumption of individual constituents results in a constant rate of heat dissipation. Then, in contrast with the control of passive systems, dissipation now stems from two sources: (i) the perturbation by an external operator, and (ii) the internal energy consumption of particles. For a sufficiently large protocol duration, where the dissipation from the perturbation is small, the contribution from internal activity is ever-growing. For a sufficiently small protocol duration, the fast protocol incurs a large dissipation from the large perturbation, as with passive systems. As a result, slow protocols are no longer optimal for the active case, and, now, a finite protocol duration minimizes the dissipation [Fig.~\ref{fig:Sketch}]. Therefore, the control of active systems involves not only finding optimal trajectories at fixed duration, but also finding which duration achieves the best trade-off between internal and external dissipation. An important challenge is then to rationalize how the optimal duration depends on the interplay between the external perturbation and the internal drive.

In this paper, inspired by previous efforts on passive systems, we herein utilize stochastic thermodynamics and response theory to systematically derive an optimal thermodynamic control framework for active systems. In so doing we build upon recent developments in and applications of linear response theory for active systems~\cite{DalCengio2019, Maes2020, Martin2021, DalCengio2021, Caprini2021} and nonlinear response theory for driven thermal systems~\cite{Lippiello2008B,Lippiello2008E, Basu2015a, Helden2016, Basu2018, Holsten2021}. First, we present, in Sec.~\ref{sec:control}, our theoretical framework for the control of both continuous and discrete-state active systems. Second, we show, in Sec.~\ref{sec:app}, how our framework can be deployed to inform the control of some specific model systems. We consider two cases: (i)~an active Ornstein-Uhlenbeck particle confined in a harmonic trap with varying stiffness, and (ii)~an assembly of active Brownian particles with purely repulsive interactions whose size is varied. In both cases, we obtain the optimal protocol for the corresponding control parameter, and we discuss shared implications that the presence of self-propulsion has on deriving the optimal protocols. Overall, our results provide a systematic framework to address the control of active systems, and illustrate the main differences with respect to the control of passive systems.

% ==========================================================================================

\section{Optimal control}
\label{sec:control}
We propose a systematic recipe for optimizing the control of active systems. It consists in predicting how an external operator should vary the parameters of the potential energy to drive the system between two states at minimal cost. We consider the cost function to be the heat dissipated by the system to a thermal reservoir. Assuming weak and slow driving, we decompose the heat into specific correlations and averages, whose dependence on the control parameter determines the optimal protocol. Interestingly, the heat always features a minimum at the protocol time which achieves the best trade-off between internal and external dissipation [Fig.~\ref{fig:Sketch}].

% ------------------------------------------------------------------------------------------

\subsection{Thermodynamic cost function}
\label{sec:cost}

We consider a system of active particles immersed in a thermal bath at temperature $T$, where each particle $i = \lbrace 1, \ldots, N \rbrace$ has an independent self-propulsion velocity $\mathbf{v}_i$ that does not depend on position nor on any details of the protocol. We describe the motion of particles by an overdamped Langevin equation given as
\begin{equation}\label{Con:eq:sec1:Langevin}
    \dot{\mathbf{r}}_i = \mu {\bf f}_i + \sqrt{2D} \pmb{\eta}_i, 
    \quad
    {\bf f}_i = - \nabla_i \phi(\alpha, \lbrace \mathbf{r} \rbrace) + \mathbf{v}_i/\mu ,
\end{equation}
where $\mathbf{r}_i$ is the position of particle $i$, $\mu$ is the mobility, $\phi(\alpha, \lbrace \mathbf{r} \rbrace)$ is the total potential energy that depends on the control parameter $\alpha$ and the set of all particle positions $\lbrace \mathbf{r} \rbrace$, $D = \mu T$ is the diffusion coefficient (with Boltzmann constant $k_\text{B} = 1$), and $\pmb{\eta}_i$ is a Gaussian white noise with zero mean and unit variance. The potential energy can account for both particle interactions and external fields. In all what follows, the protocol consists in varying $\alpha(t)$ from its initial value $\alpha_0=\alpha(t=0)$ to its final value $\alpha_1=\alpha(t=t_p)$ within the duration $t_p$.

For our control problem, we choose the cost function to be the average heat dissipated into the thermal bath (held at temperature $T$)~\cite{Sekimoto1998, Seifert2012, Tociu2019, Bo2019, Nemoto2020}, defined as
\begin{equation} \label{Con:eq:sec1:HeatDef}
    \langle Q \rangle = \frac{1}{\mu} \int_{0}^{t_p} dt \left \langle  \dot{\mathbf{r}}_i \cdot \left(\dot{\mathbf{r}}_i - \sqrt{2D} \pmb{\eta}_i \right) \right \rangle,
\end{equation}
where, unless stated otherwise, we perform a sum over repeated indices throughout this paper. Hereafter, the dot product ($\cdot$) is interpreted within Stratonovich convention, for which standard rules of differential calculus carry over to stochastic variables~\cite{Gardiner}. Substituting the dynamics from Eq.~\eqref{Con:eq:sec1:Langevin} into Eq.~\eqref{Con:eq:sec1:HeatDef}, and using the chain rule $\dot{\mathbf{r}}_i \cdot \nabla_i \phi = \dot{\phi} - \dot{ \alpha} \partial_\alpha \phi$, we obtain
\begin{equation}\label{equ:heatraw}
	\langle Q \rangle = \langle \phi_0 \srangle - \langle \phi_1 \rangle + \int_{0}^{t_p} dt \Big[ \dot{ \alpha}\left \langle \partial_\alpha \phi \right \rangle + \langle J \rangle \Big] ,
\end{equation}
where $\langle X \srangle$ denotes the steady-state average of observable $X$, $\phi_0$ and $\phi_1$ are the initial and final potential energies respectively, and
\begin{equation}
	J = {\bf f}_i \cdot {\bf v}_i.
\end{equation}
We have assumed that the system is in a steady state at $t=0$, yet it need not be the case at $t=t_p$. Equation~\eqref{equ:heatraw} can be regarded as the extension of the first law of thermodynamics (namely, the conservation of energy) for active systems~\cite{FodorCates2021, fodor2021irreversibility}. In addition to the potential change $\langle \phi_0 \srangle - \langle \phi_1 \rangle$ and the work rate $\dot{ \alpha}\left \langle \partial_\alpha \phi \right \rangle$, that are also present for passive systems, the heat features the dissipation rate $J$. In the passive limit (${\bf v}_i={\bm 0}$ for all $i$) $J$ vanishes, and one recovers the first law of thermodynamics in its standard form.

The self-propulsion term ${\bf v}_i/\mu$ in Eq.~\eqref{Con:eq:sec1:Langevin} describes the force on particle $i$ that results from the conversion of energy into directed motion. In active systems such a conversion involves multiple degrees of freedom, deliberately discarded in our framework, that typically provide additional contributions to the heat. Other models of active matter have been proposed which resolve these underlying nonequilibrium processes~\cite{Pietzonka2017,Speck2018,Gaspard2019, Gaspard2020,Speck2022}. Interestingly, for the case where the chemical reactions between fuel molecules are always transduced into the motion of the active particle, any extra dissipation due to the presence of chemical reactions is only a background contribution, independent of the potential $\phi$. Therefore, all the relevant contributions to the heat arising from $\phi$ are already accounted for in Eq.~\eqref{equ:heatraw}.

In the absence of external control ($\dot\alpha=0$), the heat reduces to the last term in Eq.~\eqref{equ:heatraw}, which scales linearly with time: $\langle Q\rangle = \langle J \srangle t_p$. Such a scaling illustrates that an active system, even when at rest, dissipates heat at a constant rate in order to sustain the self-propulsion of the particles. For a quasi-static protocol, namely for protocol durations much greater than the slowest relaxation timescale of the system ($t_p \gg \tau_{\text{max}}$), $\alpha$ will change more slowly than any relaxation timescale of the system. Thus, the system goes through a series of steady states, so that all averages in Eq.~\eqref{equ:heatraw} are now steady state averages:
\begin{equation}\label{eq:fig1equ}
	\langle Q\rangle \underset{t_p \gg \tau_{\text{max}}}{=} \langle \phi_0 \srangle - \langle \phi_1 \srangle + \int_{\alpha_0}^{\alpha_1} \langle\partial_\alpha\phi\srangle d\alpha + \langle J \srangle t_p .
\end{equation}
At large times, the heat scales linearly with protocol duration [Fig.~\ref{fig:Sketch}].

We are interested in describing corrections to the quasi-static heat that permit the prediction of optimal control protocols. To this end, we focus on cases where $\alpha$ varies weakly and slowly throughout the protocol. Thus, we can then express the average of a given observable $\langle X \rangle$ in terms of the linear ($R_1$) and second-order ($R_2$) response functions as
\begin{equation}\label{eq:resp}
	\begin{aligned}
		\langle X(t)\rangle &= \langle X \srangle + \int_{-\infty}^t dt'\Delta\alpha_{t,t'} R_1(X;t,t') 
		\\
		&\quad + \int_{-\infty}^t dt' \int_{-\infty}^{t'}  dt'' \Delta\alpha_{t,t'} \Delta\alpha_{t,t''} R_2(X;t,t',t'')
		\\
		&\quad+ {\cal O}(\Delta\alpha^3) ,
	\end{aligned}
\end{equation}
with the time ordering $t \geq t' \geq t''$, and where $\Delta\alpha_{t,t'} \equiv \alpha(t)-\alpha(t')$ is the external perturbation to the control parameter. The response functions will be explicitly defined later in Eq.~\eqref{eq:resp_b}. We then expand the perturbation as
\begin{equation}\label{eq:increm}
	\Delta\alpha_{t,t'} = \dot\alpha(t) \Delta t + \frac{{\Delta t}^2}{2} \ddot\alpha(t) + {\cal O}({\Delta t}^3),
\end{equation}
where $\Delta t=t-t'$ is the time increment. The expansions in Eqs.~(\ref{eq:resp}-\ref{eq:increm}) are inspired by a previous work on the control of passive systems~\cite{Sivak2012}, although, here, we expand to higher order in both $\Delta\alpha$ and $\Delta t$. In all what follows, we assume that these expansions are valid at all times within the protocol duration. It amounts to neglecting any abrupt change in the trajectory of $\alpha$, and also regarding the averages in Eq.~\eqref{equ:heatraw} as smooth functions of time $t$.

Within this setting, we show in Appendix~\ref{Appendix:heatinresponse} that the heat can be cast as
\begin{equation}\label{Con:eq:sec1:mainHeat}
	\langle Q \rangle = B(\alpha_0,\alpha_1,\dot{\alpha}_0, \dot{\alpha}_1) + \int_0^{t_p} dt \mathcal{L}(\dot{\alpha}, \alpha),
\end{equation}
where $\dot\alpha_0 = \dot\alpha(t=0)$ and $\dot\alpha_1 = \dot\alpha(t=t_p)$, and we have introduced the boundary term
\begin{equation}\label{eq:bound}
	\begin{aligned}
		B &= \langle \phi_0 \srangle -  \langle \phi_1 \srangle  + \dot{\alpha}_0\Sigma(\alpha_0) - \dot{\alpha}_1 (\Phi+\Sigma)(\alpha_1) 
    \\
    &\quad  + \int_{\alpha_0}^{\alpha_1} d\alpha \Lambda(\alpha) ,
	\end{aligned}
\end{equation}
and the Lagrangian
\begin{equation}\label{eq:lag}
	\mathcal{L}(\dot{\alpha},\alpha) = \dot{\alpha}^2 (\Psi - \Sigma')(\alpha) - V(\alpha),
	\quad
	\Sigma' = d\Sigma/d\alpha .
\end{equation}
The functions $\{ V, \Phi, \Sigma, \Lambda, \Psi\}$ are given by
\begin{equation}\label{equ:Greeksmain}
	\begin{aligned}
  	V &= - \langle J \srangle,
		\qquad
	    \Phi = \int_{0}^{\infty} dt R_1({\phi};t,0)t,
		\\
		\Sigma &= \frac{1}{2} \int_{0}^\infty dt R_1\left(J;t,0\right) t^2,
		\\
		\Lambda &= \langle \partial_\alpha \phi \srangle + \int_{0}^{\infty} dt R_1\left(J ; t,0\right)t,
		\\
		\Psi &= \int_{0}^\infty dt R_1({\partial_\alpha \phi};t,0) t
		\\
		&\quad + \frac{1}{2} \int_{0}^\infty \int_{0}^{\infty} dt dt' R_2\left(J;t',t'-t,0\right)tt',
	\end{aligned}
\end{equation}
The decomposition of heat in Eqs.~(\ref{Con:eq:sec1:mainHeat}-\ref{equ:Greeksmain}) is one of the central results of this paper. By measuring response functions and averages at different values of $\alpha$, one can systematically construct the dependence of the Lagrangian $\cal L$ in terms of $\alpha$ and $\dot\alpha$, and deduce the expression of the corresponding heat for any protocol. Indeed, experimentally measured linear and nonlinear response functions~\cite{Molaei2023} can be injected into Eqs.~(\ref{Con:eq:sec1:mainHeat}-\ref{equ:Greeksmain}) to explicitly determine the functional dependence of the heat on the control parameter. The optimal protocol readily follows from $\cal L$ by straightforward optimization [Sec.~\ref{sec:prot}]. Importantly, Eqs.~(\ref{Con:eq:sec1:mainHeat}-\ref{equ:Greeksmain}) hold for any potential $\phi$ and self-propulsion ${\bf v}_i$, which highlights the versatility of our approach.

In the absence of self-propulsion (${\bf v}_i={\bf 0}$), the Lagrangian reduces to ${\cal L} = \dot\alpha^2 \Psi(\alpha)$, in agreement with previous results on controlling passive systems~\cite{Sivak2012}. As detailed in Sec.~\ref{sec:prot}, the difference in $\cal L$ between active and passive systems drastically affects the corresponding dependence of $\langle Q\rangle$ in terms of $t_p$. The heat monotonically decreases in the passive case, whereas it features a global minimum in the active case [Fig.~\ref{fig:Sketch}]. Therefore, although our approach only accounts for leading-order corrections to the quasistatic regime, it is actually sufficient to capture the main qualitative change between optimizing either passive or active matter.

% ------------------------------------------------------------------------------------------

\subsection{Optimal protocol}
\label{sec:prot}

Our aim is to demonstrate that the heat decomposition in Eqs.~(\ref{Con:eq:sec1:mainHeat}-\ref{equ:Greeksmain}) entails a series of generic properties, both for the optimal protocol $\alpha(t)$ and for the corresponding optimal heat $\langle Q\rangle$. Interestingly, we derive such properties without actually specifying the explicit dependence of the functions $\{ V, \Phi, \Sigma, \Lambda, \Psi\}$ on $\alpha$ [Eq.~\eqref{equ:Greeksmain}]. For an arbitrary potential $\phi$ and self-propulsion $ {\bf v}_i$, we then predict how $\langle Q\rangle$ scales with the protocol time $t_p$, and we provide an approximate estimation of the value $t_p^*$ minimizing $\langle Q\rangle$.

Our derivations essentially rely on drawing an analogy between the Lagrangian $\cal L$ in Eq.~\eqref{eq:lag}, and that of a harmonic oscillator with position-dependent mass~\cite{Khlevniuk2018}. In that respect, $V(\alpha)$ and $\dot\alpha^2 (\Psi-\Sigma')(\alpha)$ respectively stand for the potential and kinetic energies, where the effective mass $\Psi-\Sigma'$ here depends on the control parameter $\alpha$. In what follows, we assume that $V(\alpha)$ is bounded and $(\Psi-\Sigma')(\alpha)\geq0$ for all $\alpha$ between $\alpha_0$ and $\alpha_1$, which ensures the validity of our analogy. The optimal trajectory then obeys the following Euler-Lagrange equation:
\begin{equation}\label{eq:Eul}
	2 \ddot{\alpha} (\Psi- \Sigma')(\alpha) + \dot{\alpha}^2 (\Psi' - \Sigma'')(\alpha) = - V'(\alpha) .
\end{equation}
As in standard Hamiltonian mechanics, one can also display the optimal trajectory in terms of a first integral of motion. Multiplying both sides of Eq.~\eqref{eq:Eul} by $\dot\alpha$, and integrating with respect to time $t$, we get
\begin{equation}\label{equ:1stIntegral}
	E = \dot{\alpha}^2 (\Psi - \Sigma')(\alpha) + V(\alpha) .
\end{equation}
The term $E$ is analogous to the total energy in Hamiltonian mechanics. It is constant throughout the protocol, and it is constrained by $E > \underset{\alpha}{\max} \,V$.

We deduce from Eq.~\eqref{equ:1stIntegral} a relation between the protocol trajectory $\alpha(t)$ and the protocol speed $\dot\alpha(t)$:
\begin{equation}\label{equ:controlspeed}
    \dot{\alpha} = \pm \sqrt{ \frac{E-V(\alpha)}{(\Psi - \Sigma')(\alpha)} } ,
\end{equation}
which shows that the optimal trajectory obeys a first-order ordinary differential equation. This differential equation can be solved by separation of variables, and it is fully determined by $\{E,V,\Psi,\Sigma\}$. In practice, the constant of motion $E$ implicitly depends on the initial and final values $\{\alpha_0,\alpha_1\}$, and on the protocol duration $t_p$. Indeed, separating variables in Eq.~\eqref{equ:controlspeed}, and integrating throughout the protocol, we get
\begin{equation}\label{equ:tp}
	t_p = \left| \int_{\alpha_0}^{\alpha_1} d\alpha \sqrt{\frac{(\Psi - \Sigma')(\alpha)}{E - V(\alpha)}} \right| .
\end{equation}
Therefore, combining Eqs.~(\ref{equ:controlspeed}-\ref{equ:tp}), the solution of the optimal trajectory $\alpha(t)$ follows directly. Although its explicit expression cannot be given analytically for generic potential $\phi$ and self-propulsion ${\bf v}_i$, numerical integration readily yields $\alpha(t)$ for given expressions of $\{V,\Psi,\Sigma\}$, and for arbitrary $\{\alpha_0,\alpha_1\}$ and $t_p$.

Interestingly, we demonstrate in Appendix~\ref{Appendix:scalingtp} that Eqs.~(\ref{equ:controlspeed}-\ref{equ:tp}) are actually sufficient to predict how $\langle Q\rangle$ scales with protocol duration $t_p$:
\begin{equation}\label{eq:heat_scaling}
	\begin{aligned}
		&\langle Q\rangle \underset{t_p \ll t_p^*}{\sim} K(\alpha_0, \alpha_1) / t_p ,
		\\
		&\langle Q\rangle \underset{t_p \gg t_p^*}{\sim} t_p \,\underset{\alpha}{\min} \langle J\srangle ,
	\end{aligned}
\end{equation}
where the expression of $K(\alpha_0,\alpha_1)$ follows from that of $\{\Phi,\Psi,\Sigma\}$ [Eq.~\eqref{equ:Ka0a1}]. These scalings correspond to regimes where the dissipation is dominated either by the external driving of the control parameter $\alpha$ (namely for $t_p\ll t_p^*$), or by the internal self-propulsion ${\bf v}_i$ (namely for $t_p\gg t_p^*$). The former reproduces the scaling expected for passive systems~\cite{Sivak2012}. The latter agrees with the quasistatic prediction [Eq.~\eqref{eq:fig1equ}], where here the optimal protocol achieves the minimum value of $\langle J\rangle_s$, as expected.

In between the regimes of large and small $t_p$, the heat reaches a minimum at $t_p=t_p^*$ [Fig.~\ref{fig:Sketch}]. The protocol duration $t_p^*$ achieves the best compromise between two regimes of high $\langle Q\rangle$.  By extending the scalings in Eq.~\eqref{eq:heat_scaling} to regimes where $t_p\approx t_p^*$, we approximate $t_p^*$ as the value of $t_p$ where these scalings have same $\langle Q\rangle$:
\begin{equation}\label{equ:crudetp}
	t_p^\ast \approx \sqrt{ K(\alpha_0,\alpha_1) \big/ \underset{\alpha}{\min} \langle J\srangle} .
\end{equation}
Equation~\eqref{equ:crudetp} illustrates that $t_p^*$ sets a trade-off between (i) the heat due to external drive, involving the system relaxation through response functions in the definition of $K(\alpha_0,\alpha_1)$ (equivalently in the definition of $\{\Phi,\Psi,\Sigma\}$, see Eq.~\eqref{equ:Greeksmain}), and (ii) the heat due to internal self-propulsion, involving the steady state average $\langle J\rangle_s$. A qualitatively similar trade-off was reported in a different context for systems combining autonomous cycles and external drives~\cite{Machta2020}.

In the passive limit (${\bf v}_i={\bf 0}$), we deduce from Eq.~\eqref{equ:crudetp} that $t_p^*$ diverges. The heat now reaches its minimum asymptotically at large $t_p$ [Fig.~\ref{fig:Sketch}], and the scaling corresponding to $t_p\ll t_p^*$ in Eq.~\eqref{eq:heat_scaling} now extends to all $t_p$. Indeed, as expected from the thermodynamics of passive systems, the protocol with smallest dissipation is always the slowest, in which case the heat reduces to its quasistatic value.

Interestingly, when controlling active systems, the boundary term $B$ in Eq.~\eqref{eq:bound} involves the protocol speeds $\dot\alpha$ at initial and final times, respectively $\dot\alpha_0$ and $\dot\alpha_1$. Within our framework, optimizing the Lagragian term $\cal L$ already leads to fixing $\dot\alpha_0$ and $\dot\alpha_1$ [Eqs.~(\ref{equ:controlspeed}-\ref{equ:tp})] for a given choice of $\{\alpha_0,\alpha_1\}$ and $t_p$. To optimize simultaneously $B$ and $\cal L$, one could potentially consider abrupt changes in the protocol trajectory $\alpha(t)$ at initial and final times~\cite{Schmiedl2007}. Yet, such discontinuous jumps would not be consistent with the expansions in Eqs.~(\ref{eq:resp}-\ref{eq:increm}). Therefore, in what follows, we focus on the optimal trajectory given by Eqs.~(\ref{equ:controlspeed}-\ref{equ:tp}).

% ------------------------------------------------------------------------------------------

\subsection{Response functions}
\label{sec:responses}

We now demonstrate that the response functions in the heat decomposition of Eqs.~(\ref{Con:eq:sec1:mainHeat}-\ref{equ:Greeksmain}) can be expressed in terms of some specific correlations functions. Indeed, evaluating response functions generally requires measuring how the system is affected by weak perturbation, which can prove difficult both in numerical simulations and in experiments. Instead, our aim is to show that the response is actually already encoded in spontaneous fluctuations of the unperturbed dynamics.

In passive systems, the fluctuation-dissipation theorem enforces generic relations between linear response and correlation functions~\cite{Kubo1966}, and similar types of relations also exist for the second-order response~\cite{Basu2015, Basu2015a, Helden2016}. Interestingly, recent works have shown that response-correlation relations can be extended beyond passive systems~\cite{Maes2020}, and in particular to active matter~\cite{DalCengio2019, Martin2021, DalCengio2021, Caprini2021}. Here, we explicitly derive such relations for the active dynamics in Eq.~\eqref{Con:eq:sec1:Langevin}.

From the definition of the response functions in Eq.~\eqref{eq:resp} for a generic observable $X$, we express $R_1$ and $R_2$ as
\begin{equation}\label{eq:resp_b}
	\begin{aligned}
		R_1(X;t,t') &= \frac{\delta\langle X(t) \rangle}{\delta \Delta\alpha_{t,t'}} \bigg|_{\Delta\alpha\to0} ,
		\\
		R_2(X;t,t',t'') &= \frac{\delta^2\langle X(t) \rangle}{\delta \Delta\alpha_{t,t'}\delta \Delta\alpha_{t,t''}} \bigg|_{\Delta\alpha\to0} .
	\end{aligned}
\end{equation}
Evaluating the response functions then amounts to predicting how the average $\langle X(t)\rangle$ is perturbed when varying the control parameter $\alpha$. To this end, we express this average in terms of a path integral as
\begin{equation}\label{eq:path}
	\begin{aligned}
		\langle X(t)\rangle &= \int X(t) \prod_{i} P_\eta[{\bm\eta}_i] P_v[{\bf v}_i] {\cal D}{\bm\eta}_i {\cal D}{\bf v}_i
		\\
		&\quad\times \delta\big( \dot{\bf r}_i - \mu {\bf f}_i({\bf r}_i,{\bf v}_i) + \sqrt{2D} {\bm\eta}_i \big) {\cal D}{\bf r}_i ,
	\end{aligned}
\end{equation}
where $P_\eta$ and $P_v$ denote the probabilites to observe a given trajectory for the noise terms ${\bm\eta}_i$ and ${\bf v}_i$, respectively. We have used that these noises are independent. Using that $P_\eta$ is Gaussian, integrating Eq.~\eqref{eq:path} with respect to position trajectories $\{{\bf r}_i\}$ yields
\begin{equation}\label{eq:path_b}
	\langle X(t)\rangle = {\cal N} \int X(t) \,e^{-{\cal S}} {\cal D}[{\bm\eta}, {\bf v}] ,
\end{equation}
where ${\cal D} [{\bm\eta}, {\bf v}] = \prod_i P_v[{\bf v}_i] {\cal D}{\bm\eta}_i {\cal D}{\bf v}_i$, and the normalization factor is given by $1/{\cal N} = \int e^{-{\cal S}} {\cal D} [{\bm\eta}, {\bf v}]$. The term $\cal S$ is the standard Onsager-Machlup action~\cite{Onsager1953}:
\begin{equation}
	\mathcal{S} = \frac{1}{4D} \int_0^{t_p} dt \left( \dot{\mathbf{r}}_i - \mu\mathbf{f}_i \right)^2 .
\end{equation}
Note that, although the dynamics is given in Stratonovich convention throughout the paper, $\cal S$ does not feature the term $(\mu/2)\int dt \nabla_i\cdot{\bf f}_i$ usually given in the Onsager-Machlup action. This is because the path integral in Eq.~\eqref{eq:path_b} is written with respect to noise trajectories, instead of position trajectories.

Substituting the path integral of Eq.~\eqref{eq:path_b} into Eq.~\eqref{eq:resp_b}, we deduce
\begin{equation}\label{eq:path_c}
	\begin{aligned}
		R_1(X;t,t') &= \int X(t) \frac{\delta\big({\cal N} e^{-{\cal S}}\big)}{\delta \Delta\alpha_{t,t'}} \bigg|_{\Delta\alpha\to0} {\cal D} [{\bm\eta}, {\bf v}] ,
		\\
		R_2(X;t,t',t'') &= \int X(t) \frac{\delta^2\big({\cal N} e^{-{\cal S}}\big)}{\delta \Delta\alpha_{t,t'}\delta \Delta\alpha_{t,t''}} \bigg|_{\Delta\alpha\to0} {\cal D} [{\bm\eta}, {\bf v}] ,
	\end{aligned}
\end{equation}
where we have used that ${\cal D} [{\bm\eta}, {\bf v}]$ is independent of $\alpha$, since noise trajectories are not affected by the control parameter. Then, predicting how $\cal S$ varies with $\alpha$ is a route to expressing $R_1$ and $R_2$ in terms of correlation functions. The perturbation $\alpha\to\alpha-\Delta\alpha$ affects the action as $\mathcal{S}\to\mathcal{S} + \Delta\mathcal{S}$, where
\begin{equation}\label{eq:action}
	\begin{aligned}
		 \Delta \mathcal{S} &= \frac{1}{4T} \int_0^{t_p} dt  \Big[ - 2 \Delta\alpha (\nabla_i \partial_\alpha \phi) \cdot \left( \dot{\mathbf{r}}_i - \mu \mathbf{f}_i \right)
		\\
		&\quad+ \Delta\alpha^2 \left( \mu (\nabla_i \partial_\alpha \phi)^2 + (\nabla_i \partial_\alpha^2 \phi)\cdot ( \dot{\mathbf{r}}_i - \mu \mathbf{f}_i) \right)\Big] 
  \\&\quad + \mathcal{O}(\Delta \alpha^3).
	\end{aligned}
\end{equation}
As detailed in Appendix~\ref{app:ccos}, combining Eqs.~(\ref{eq:path_c}-\ref{eq:action}) then leads to
\begin{equation}\label{equ:R1cont}
	2T R_1(X;t,t') = \frac{d}{dt'} \llangle[\big]{X}_t [\partial_\alpha \phi]_{t'} \rrangle[\big] - \mu \llangle[\big] X_t \big[ (\nabla_i \partial_\alpha \phi)\cdot \mathbf{f}_i \big]_{t'} \rrangle[\big].
\end{equation}
The connected correlation $\llangle X(t)Y(t') \rrangle \equiv \llangle X_t Y_{t'} \rrangle$ is defined for arbitrary observables $X$ and $Y$ as
\begin{equation}\label{eq:2pt}
	\llangle X_tY_{t'} \rrangle = \langle X_t Y_{t'} \rangle - \langle X \srangle \langle Y \srangle.
\end{equation}
In the absence of self-propulsion (${\bf v}_i={\bf 0}$), we recover the celebrated fluctuation-dissipation theorem: $R_1(X;t,t') = (1/T) (d/dt') \llangle X(t) \partial_\alpha \phi (t') \rrangle$ for $t>t'$~\cite{Kubo1966}. Indeed, this theorem is readily obtained from Eq.~\eqref{equ:R1cont} by considering the time-symmetrized response $R_1(X;t,t')-R_1(X;t',t)$, and using the time-reversal symmetry of equilibrium correlations~\cite{Maes2020} (see Appendix~\ref{app:ccos}). Similarly, we show in Appendix~\ref{app:ccos} that the second-order response can be written as
\begin{widetext}
\begin{equation} \label{equ:R2cont}
	\begin{aligned}
		&4T^2 \,R_2(X;t,t',t'') 
  \\&=  \frac{d}{dt'} \frac{d}{dt''} \lllangle[\big] {X}_t [\partial_\alpha \phi] {}_{t'} [\partial_\alpha \phi] {}_{t''} \rrrangle[\big] + \mu^2 \lllangle[\big] {X}_t \big[ (\nabla_i \partial_\alpha \phi)\cdot \mathbf{f}_i \big] {}_{t'} \big[  (\nabla_j \partial_\alpha \phi)\cdot\mathbf{f}_j \big]{}_{t''} \rrrangle[\big]
		- \mu  \frac{d}{dt''} \lllangle[\big] {X}_t [\partial_\alpha \phi] {}_{t''} \big[ (\nabla_i \partial_\alpha \phi)\cdot \mathbf{f}_i \big] {}_{t'} \rrrangle[\big] 
  \\
    &\quad - \mu  \frac{d}{dt'} \lllangle[\big] {X}_t [\partial_\alpha \phi] {}_{t'} \big[ (\nabla_i \partial_\alpha \phi)\cdot\mathbf{f}_i \big]{}_{t''} \rrrangle[\big] 
 - 2T\delta(t' - t'') \bigg[ \frac{d}{dt'} \llangle[\big] X_t [\partial_\alpha^2 \phi]{}_{t'} \rrangle[\big] + \mu  \llangle[\big] {X}_t \big[ (\nabla_i \partial_\alpha \phi)^2 - (\nabla_i \partial_\alpha^2 \phi)\cdot \mathbf{f}_i \big] {}_{t'}\rrangle[\big] \bigg] ,
	\end{aligned}
\end{equation}
\end{widetext}
where the connected correlation $\lllangle X(t)Y(t') Z(t'') \rrrangle \equiv \lllangle X_t Y_{t'} Z_{t''} \rrrangle$ is defined for arbitrary observables $X$, $Y$, and $Z$ as
\begin{equation}\label{eq:3pt}
	\begin{aligned}
		\lllangle X_t Y_{t'} Z_{t''} \rrrangle&= \langle X_{t} Y_{t'} Z_{t''} \rangle 
	     - \langle X \srangle \langle Y_{t'} Z_{t''} \rangle 
		\\&\quad - \langle Y \srangle \langle X_t Z_{t''} \rangle 
		- \langle Z \srangle \langle X_t Y_{t'} \rangle
		\\&\quad+ 2 \langle X \srangle \langle Y \srangle \langle Z \srangle.
	\end{aligned}
\end{equation}
The relations in Eqs.~(\ref{equ:R1cont}-\ref{eq:3pt}), coupled with Eqs.~(\ref{Con:eq:sec1:mainHeat}-\ref{equ:Greeksmain}), are the main relations of this paper. Indeed, they provide explicit connections between the response functions $R_1$ and $R_2$, defined for an arbitrary observable $X$ [Eq.~\eqref{eq:resp}], and correlation functions in the unperturbed dynamics. Similar linear and nonlinear response relations have been previously derived in the context of general Markov and non-equilibrium systems, with specific application to the Langevin equation and Ising spin variables ~\cite{Lippiello2008B,Lippiello2008E}. Our response relations Eqs.~(\ref{equ:R1cont}-\ref{eq:3pt}) provide a route to evaluating response functions resulting from external control without actually perturbing the system. Overall, we are now in a position to deploy these relations to the specific observables which define the response functions of interest in Eq.~\eqref{equ:Greeksmain}.

% ------------------------------------------------------------------------------------------

\subsection{Discrete-state dynamics}

In this section, we show that our generic framework for controlling the potential in Eq.~\eqref{Con:eq:sec1:Langevin}, corresponding to a continuous-state dynamics, can be extended to nonequilibrium systems with discrete states. To this end, we consider a generic dynamics that is governed by the following master equation:
\begin{equation}\label{equ:Compactmaster}
	\dot{p}_i =  \sum_j K_{ij}(\alpha) \,p_j ,
\end{equation}
where $p_i(t)$ is the probability for the system to be in the state $i$ at time $t$, and $K_{ij}$ is the transition rate to go from state $j$ to $i$. The control parameter $\alpha$ determines the expression of $K_{ij}$, analogously to how $\alpha$ controls the potential $\phi$ in Eq.~\eqref{Con:eq:sec1:Langevin}. The average of a given observable $X$ and its time derivative are expressed as
\begin{equation}\label{eq:avg}
	\begin{aligned}
		\langle X \rangle &= \sum_i X_i p_i ,
		\\
		\langle \dot X \rangle &= \sum_i \Big[ p_i \dot\alpha \partial_\alpha X_i + X_i \sum_j K_{ij} p_j \Big] ,
	\end{aligned}
\end{equation}
where $X_i$ is the projection of $X$ over the state $i$. As for the continuous-state case, the system is in contact with a heat bath at temperature $T$, and subject to active driving. In practice, we consider transition rates with Arrhenius form, ensuring that the dynamics is thermodynamically consistent~\cite{Arrhenius1889, Herpich2018}. Such a transition rate is given by
\begin{equation} \label{equ:Rates}
	K_{ij}(\alpha) = c_{ij} \exp\left(- \frac{\phi_j(\alpha) - \phi_i(\alpha) + \varepsilon_{ij}}{2T} \right)
\end{equation}
where $c_{ij}=c_{ji}$ is the rate amplitude, $\phi_i$ is the potential energy of state $i$, and $\varepsilon_{ij}= - \varepsilon_{ji}$ is the energy exchange associated with the active driving.

The heat dissipated into the bath while varying $\alpha$ from $\alpha_0$ to $\alpha_1$, within a time duration $t_p$, reads~\cite{Lebowitz1999, VandenBroeck2015}
\begin{equation}\label{eq:heat_disc}
	\langle Q \rangle = T \int_0^{t_p} dt \sum_{i,j} K_{ij} p_j \ln \frac{K_{ij}}{K_{ji}} .
\end{equation}
Substituting the definition of the transition rates from Eq.~\eqref{equ:Rates} into Eq.~\eqref{eq:heat_disc}, we get
\begin{equation}\label{equ:discreteHeat}
	\langle Q \rangle = \langle \phi(\alpha_0) \srangle - \langle \phi(\alpha_1) \rangle + \int_0^{t_p} dt \Big[ \dot{\alpha} \langle \partial_\alpha \phi \rangle + \langle J\rangle \Big],
\end{equation}
where we have used the definition of averages given in Eq.~\eqref{eq:avg}, and we have introduced the active dissipation rate:
\begin{equation}
	\langle J\rangle = \sum_{i,j} K_{ij} p_j \varepsilon_{ij} .
\end{equation}
The first law of thermodynamics in Eq.~\eqref{equ:discreteHeat} takes a similar form as for the continuous-state case in Eq.~\eqref{equ:heatraw}. Therefore, within the same set of assumptions as in Sec.~\ref{sec:cost}, the decomposition of heat as the sum of boundary and Lagrangian term [Eqs.~(\ref{Con:eq:sec1:mainHeat}-\ref{equ:Greeksmain})] remains valid for the discrete-state case, as with all the results for the optimal protocol in Sec.~\ref{sec:prot}.

Again, to connect response functions to correlation functions, as for the case of continuous-state dynamics [Sec.~\ref{sec:responses}], we rely on a path-integral approach. First, we define the path weight of a given trajectory $\omega([0,t_p])$ which goes through a series of discrete states within the protocol time $t_p$~\cite{Seifert2012, Maes2007, maes2020frenesy}:
\begin{equation}
	P = \mathcal{N} \exp(-\mathcal{A}) ,
\end{equation}
with the normalization factor $\mathcal{N}$. The action $\cal A$ reads 
\begin{equation} \label{equ:DisAction}
	\mathcal{A} = \int_0^{t_p} dt \sum_{i,j\neq i} \left[  (K_{ij} - \bar{K}_{ij}) \rho_j(t) + n_{ij}(t) \ln \frac{\bar{K}_{ij}}{K_{ij}} \right] ,
\end{equation}
where we have introduced the reference transition rate $\bar{K}_{ij}$ that does not depend on $\alpha$. The stochastic density $\rho_j(t)$ is $1$ if the trajectory $\omega([0,t])$ goes through the state $j$ at time $t$, and $0$ otherwise:
\begin{equation}
	 \rho_j(t)= \delta_{\omega(t),j} ,
\end{equation}
where $\delta_{\omega,j}$ is the Kronecker delta function. The stochastic transition rate $n_{ij}$ is defined as
\begin{equation}
	n_{ij}(t) = \sum_\lambda \delta \big(t - \gamma_{ij}^{(\lambda)} \big),
\end{equation}
where $\gamma_{ij}^{(\lambda)}$ denotes the series of times where the system jumps from state $j$ to $i$, each one of them labelled by the index $\lambda$. The variation ${\cal A}\to{\cal A}+\Delta \mathcal{A}$ due to the perturbation $\alpha\to \alpha-\Delta\alpha$ then follows as
\begin{equation}
	\begin{aligned}\label{eq:dadis}
		\Delta \mathcal{A} &= \int_0^{t_p} dt \sum_{i,j\neq i} \Bigg[ -\Delta \alpha \frac{\partial_\alpha (\phi_i - \phi_j )}{2T} \left( n_{ij} - K_{ij} \rho_j \right)
		\\
		&\quad - \frac{\Delta \alpha^2}{2} \Bigg( \frac{\partial_\alpha^2 (\phi_i - \phi_j)}{2T} \left( n_{ij} - K_{ij} \rho_j \right) 
  \\&\quad - K_{ij} \rho_j \frac{(\partial_\alpha (\phi_i - \phi_j))^2}{4T^2} \Bigg) \Bigg] + \mathcal{O}(\Delta \alpha^3) .
	\end{aligned}
\end{equation}
Using the same procedure as that yielding the responses in the continuous-state case, we show in Appendix~\ref{app:diss} that the linear response $R_1$ in the discrete case can be written as
\begin{equation}\label{equ:R1disc}
    2T R_1(X;t,t') = \frac{d}{dt'}\llangle X_t [\partial_\alpha \phi]_{t'} \rrangle-\llangle X_t \left[K \partial_\alpha\phi\right]_{t'} \rrangle ,
\end{equation}
and the second-order response $R_2$ follows as
\begin{widetext}
\begin{equation}\label{equ:R2disc}
	\begin{aligned}
		&4T^2\, R_2(X;t,t',t'') 
  \\&= \frac{d}{dt'} \frac{d}{dt''} \lllangle[\big] X_t [\partial_\alpha \phi]_{t'} [\partial_\alpha \phi]_{t''} \rrrangle[\big] - \frac{d}{dt'} \lllangle[\big] X_t[\partial_\alpha \phi]_{t'} [K \partial_\alpha \phi]_{t''} \rrrangle[\big] - \frac{d}{dt''} \lllangle[\big] X_t[\partial_\alpha \phi]_{t''} [K \partial_\alpha \phi]_{t'} \rrrangle[\big] + \lllangle[\big] X_t [K \partial_\alpha \phi]_{t'} [K \partial_\alpha \phi]_{t''} \rrrangle[\big]
  \\& \quad - \delta(t'-t'') \left[ 2T \llangle[\big] X_t [K\partial_\alpha^2 \phi]_{t'} \rrangle[\big] - 2T \frac{d}{dt'} \llangle[\big] X_t[\partial_\alpha^2 \phi]_{t'} \rrangle[\big] + \llangle[\big] X_t [K(\partial_\alpha \phi)^2 - 2 (\partial_\alpha \phi) K (\partial_\alpha \phi)]_{t'}\rrangle[\big] \right] ,
		\end{aligned}
	\end{equation}
\end{widetext}
where, for a given observable $Y$, we have introduced the notations
\begin{equation}
	K Y = \sum_{i,j\neq i} K_{ij} \rho_j Y_i ,
	\quad
	YK Y = \sum_{i,j\neq i} K_{ij} \rho_j Y_j Y_i .
\end{equation}
Thus, Eqs.~(\ref{equ:R1disc},\ref{equ:R2disc}) provide explicit relations between response and correlation functions, which mirror those of the continuous-state case [Eqs.~(\ref{equ:R1cont},\ref{equ:R2cont})] (also see \cite{Lippiello2008E,Lippiello2008B}). Substituting these relations into the expressions in Eq.~\eqref{equ:Greeksmain} then allows one to determine the boundary and Lagrangian terms of the dissipated heat.

% ===========================================================================================

\section{From one- to many-body control}
\label{sec:app}
We now apply our systematic recipe to the control of continuous-state systems in specific examples. First, in Sec.~\ref{sec:harm}, we analytically derive the optimal protocol for an active particle subject to a harmonic potential whose stiffness varies. Second, in Sec.~\ref{sec:rep}, we address the case of an assembly of repulsive active Brownian particles (RABPs) with a controllable size, that undergoes motility-induced phase separation (MIPS) at high packing fraction.

% ------------------------------------------------------------------------------------------

\begin{figure}
\includegraphics[width=\columnwidth]{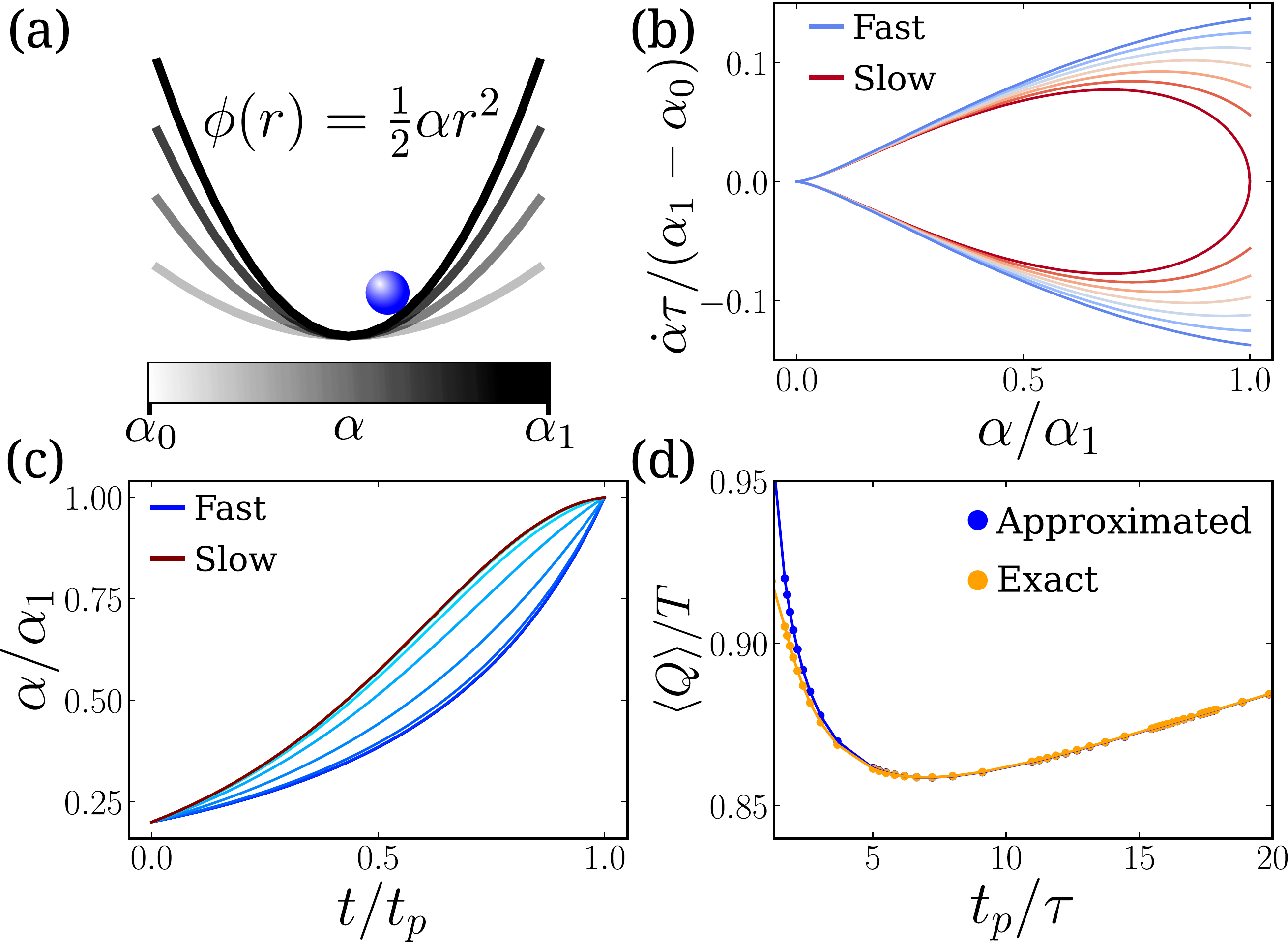}
\caption{\label{fig:1Dcase} Optimal thermodynamic control of an active particle in a one-dimensional trap. (a)~Illustration of an active particle (blue) in a harmonic trap. The protocol consists in driving the trap strength from $\alpha_0$ to $\alpha_1$. (b)~Control speed $\dot{\alpha}$ [Eq.(\ref{equ:controlspeed})] against trap strength $\alpha$ for different $t_p = \lbrace 0.1\tau, 1\tau, 5\tau, 10\tau, 20\tau, 100\tau \rbrace$. (c)~Protocols for $t_p =\lbrace 0.1\tau, \ldots, 100\tau \rbrace $. (d)~Comparing the scaled heat, where $T=D/\mu$, against protocol duration as approximated in the response framework (blue) [Eq.~\eqref{Con:eq:sec1:mainHeat}] and the exact expression (orange) [Eq.~\eqref{eq:1heatdef}]. For both curves, we use the protocols shown in (c). Parameters: $\mu=1$, $D=1$, $D_1=0.01$, $\tau = 1$, $\alpha_0 = 1$, and $\alpha_1 = 5$.}
\end{figure}

\subsection{Active particle in a harmonic trap}
\label{sec:harm}

Having established the main framework, we now test it on a toy model consisting of a single active particle in a one-dimensional harmonic trap [Fig.~\ref{fig:1Dcase}(a)]. The motion of the particle is governed by
\begin{equation} \label{eq:1dmodel}
	\begin{aligned}
		\dot{r} = \mu f + \sqrt{2D} \eta, 
    \quad
    f= - \alpha r + v/\mu, 
	\end{aligned}
\end{equation}
where $\phi(\alpha,r) = \alpha r^2/2$ is the trap potential, with $\alpha$ being the stiffness. For the self-propulsion, we choose the simplest Ornstein-Uhlenbeck process with $\langle v \rangle=0$ and $\langle v(t) v(s) \rangle = \tau^{-1} D_1 e^{-|t-s|/\tau}$, where $\tau$ is the persistence time and $D_1$ is the amplitude of the active noise. The heat [Eq.~\eqref{equ:heatraw}] associated with varying $\alpha$ from $\alpha(0)=\alpha_0$ to $\alpha(t_p)=\alpha_1$ reads
\begin{equation}\label{eq:1heatdef}
	\langle Q\rangle = \frac{\alpha_0}{2} \langle r_0^2\srangle - \frac{\alpha_1}{2} \langle r_1^2\rangle + \int_0^{t_p} dt \bigg[ \frac{\dot\alpha}{2} \langle r^2\rangle + \frac{D_1}{\tau\mu} - \alpha \langle r v\rangle \bigg],
\end{equation}
where $r_0$ and $r_1$ are respectively the initial and final positions of the particle. Since the dynamics in Eq.~\eqref{eq:1dmodel} is linear, the time evolution of the moments $\langle r^2\rangle$ and $\langle r v\rangle$ is closed, and can be obtained using It\^o's lemma~\cite{Gardiner} as
\begin{equation}\label{eq:Ito}
	\begin{aligned}
		\frac{1}{2}\frac{d}{dt} \langle r^2 \rangle &= \langle rv \rangle - \mu \alpha \langle r^2 \rangle + D,
		\\
		\tau\frac{d}{dt}{\langle rv \rangle} &= D_1 - \langle rv \rangle \left(1+ \alpha\mu  \tau\right).
	\end{aligned}
\end{equation}
In steady state, the left-hand side of Eq.~\eqref{eq:Ito} is zero, resulting in
\begin{equation} \label{equ:1DITOaves}
	\langle r^2 \srangle = \frac{1}{\alpha \mu} \left( \frac{D_1}{1+\alpha\mu\tau} + D  \right) ,
	\quad
	\langle rv \srangle = \frac{D_1}{1+\alpha\mu\tau} .
\end{equation}
Therefore, the heat given in Eq.~\eqref{eq:1heatdef} can be numerically measured for a given protocol $\alpha(t)$ simply by simulating the dynamics of the moments in Eq.~\eqref{eq:Ito}.

To evaluate the decomposition of heat [Eqs.~(\ref{Con:eq:sec1:mainHeat}-\ref{equ:Greeksmain})], we use the continuous-state response functions in Eqs.~\eqref{equ:R1cont} and~\eqref{equ:R2cont}. We show in Appendix~\ref{Appendix:1Dcase} how to obtain the following expressions
\begin{equation} \label{equ:1DGreeks}
	\begin{aligned}
		V &= \frac{\alpha D_1 }{1 + \alpha\mu\tau} - \frac{D_1}{\mu\tau} ,
		\qquad
		\Sigma = \frac{D_1  \alpha\mu\tau ^3}{(1 + \alpha\mu\tau)^4} ,
		\\
		\Phi &= \frac{D (1 + \alpha\mu\tau)^3+D_1 \big[ 1 + 3 \alpha  \mu  \tau + 4 (\alpha \mu\tau)^2 \big]}{2 (\alpha\mu)^2 (1 + \alpha\mu\tau)^3} ,
		\\
		\Lambda &= \frac{D (1 + \alpha\mu\tau)^3 + D_1 \big[1 + 2\alpha\mu\tau - (\alpha\mu\tau)^2 \big]}{2 \alpha  \mu  (1 + \alpha\mu\tau)^3} ,
		\\
		\Psi &= \frac{D}{{4\alpha^3 \mu ^2 (\alpha  \mu  \tau +1)^5}} \Big\{ (\alpha  \mu  \tau +1)^5
		\\
		&\quad + \frac{D_1}{D} \big[ 1 + 5 \alpha\mu\tau + 11 (\alpha\mu\tau)^2 + 3 (\alpha\mu\tau)^3 - 16 (\alpha\mu\tau)^4 \big]
		\\
		&\quad - \Big(\frac{D_1}{D}\Big)^2 (\alpha\mu\tau)^3 \big[ 8 + 7 \alpha\mu\tau + 4(\alpha\mu\tau)^2 + (\alpha\mu\tau)^3 \big] \Big\} .
	\end{aligned}
\end{equation}
In the passive limit ($D_1 = 0$), the expressions reduce to $\Psi = D/2 \alpha^3 \mu^2$, $\Sigma = V = 0$, and $\Lambda = \alpha \mu \Phi = D/(2 \alpha \mu)$, in agreement with~\cite{Sivak2012}. The protocols for changing the stiffness $\alpha$ of the trap can be readily obtained following the procedure detailed in Sec.~\ref{sec:prot}. First, we numerically solve the elliptic integral equation in Eq.~\eqref{equ:tp} which connects the constant $E$ with the protocol time $t_p$. As $E$ approaches $\underset{\alpha}{\max} \,V$ (namely for increasing $t_p$), one needs increasingly high precision to numerically solve Eq.~\eqref{equ:tp}. To this end, we utilize the arbitrary-precision numerical library \texttt{Arb}~\cite{Johansson2017}, that provides a state-of-the-art numerical integration procedure based upon adaptive bisection and adaptive Gaussian quadrature~\cite{Johansson2018}. Second, we obtain the protocol trajectory $\alpha(t)$ by numerically solving Eq.~\eqref{equ:controlspeed}, for given protocol time $t_p$ and boundary conditions $\{\alpha_0,\alpha_1\}$. We report the results of optimal protocols for different $t_p$ in Figs.~\ref{fig:1Dcase}(b-c). Very fast ($t_p \ll t_p^*$) and very slow ($t_p \gg t_p^*$) protocols collapse onto two separate master curves. This is in contrast with the thermodynamic control of passive systems, where there is only a single protocol master curve valid for all protocol durations~\cite{Deffner2020}.

For the optimal protocols in Fig.~\ref{fig:1Dcase}(c), we then compare two separate evaluations of heat. We either (i) simulate the dynamics in Eq.~\eqref{eq:Ito} with the optimal protocol, and measure the corresponding heat [Eq.~\eqref{eq:1heatdef}], or (ii) directly substitute the optimal protocol in the expression of heat given in Eq.~\eqref{Con:eq:sec1:mainHeat}, which relies on response theory. Comparing these two results allows us to delineate the regime where the response theory is indeed valid. We report in Fig.~\ref{fig:1Dcase}(d) that the two evaluations of heat match very well not only at large $t_p$, where the heat scales linearly with $t_p$, but also in the regime where the heat is minimum ($t_p\simeq t_p^\ast$). At times smaller than $t_p^*$, a discrepancy arises, showing that the assumption of slow driving, underpinning the response framework, breaks down in this regime.

\begin{figure}
\includegraphics[width=0.4\textwidth]{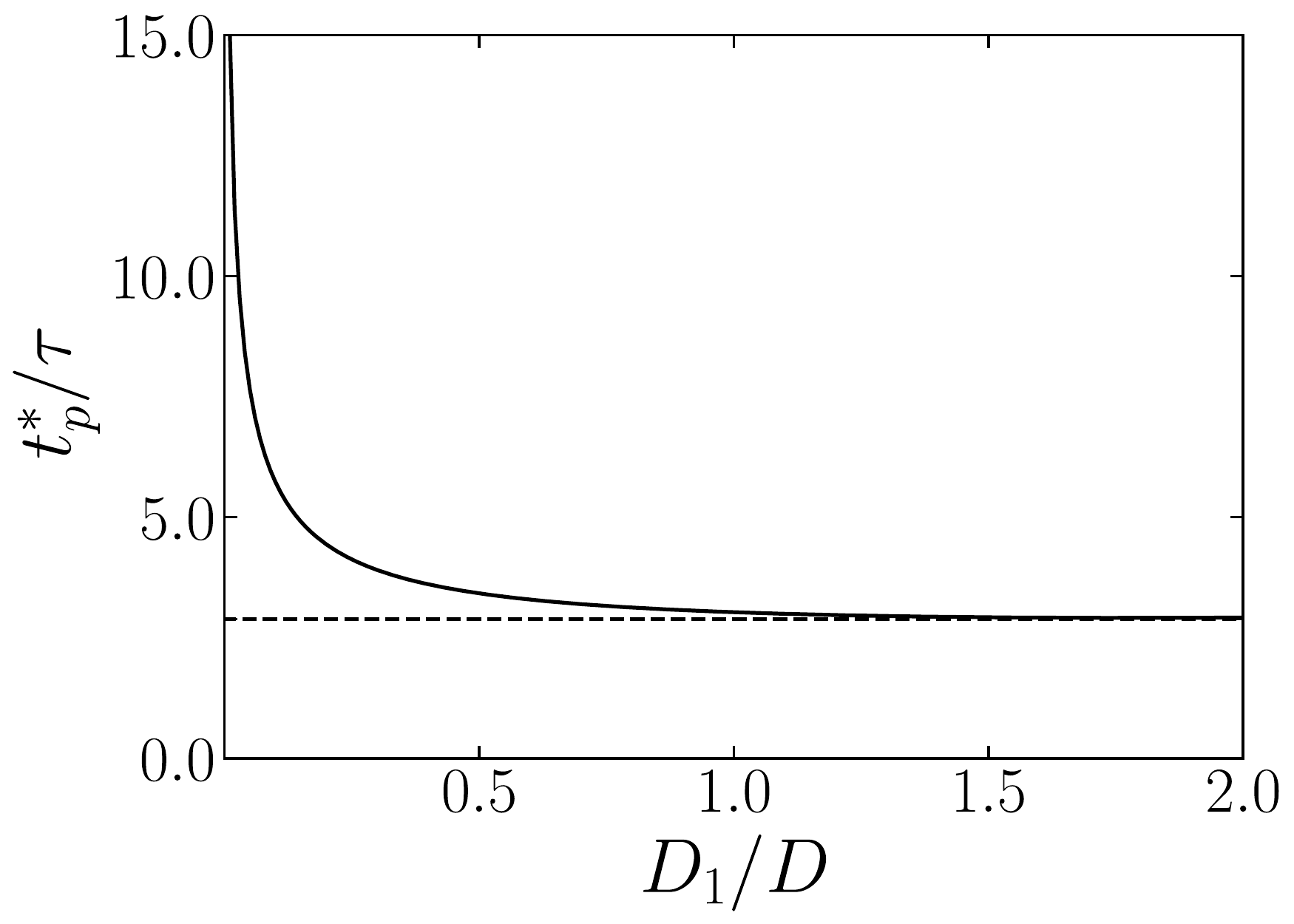}
\caption{\label{fig:tpVsD1} Optimal protocol duration $t_p^\ast$ as a function of the active noise amplitude $D_1$, as predicted by the scaling relation Eq.~\eqref{equ:crudetp} in conjunction with Eq.~\eqref{equ:1DGreeks}. Parameters: $\alpha_0/\alpha_1=1/5,\mu=\tau=D=1$.}
\end{figure}

For a general potential, we expect that the protocol time at minimum heat $t_p^*$ increases as the activity gets weaker ($|{\bf v}_i|\to 0$). For the harmonic case treated here, we evaluate analytically the approximated estimation of $t_p^*$, which stems from matching the asymptotic behaviors of the heat, by substituting Eq.~\eqref{equ:1DGreeks} into Eqs.~\eqref{equ:crudetp} and~\eqref{equ:Ka0a1}. This estimation of $t_p^*$ captures the divergence $t_p^*\sim 1/\sqrt{D_1}$ at small activity ($D_1/D\ll 1$). Interestingly, it also predicts that $t_p^*$ should plateau to a finite value at large activity ($D_1/D\gg 1$), as shown in Fig.~\ref{fig:tpVsD1}.

% ------------------------------------------------------------------------------------------

\subsection{Repulsive active Brownian particles (RABPs)}
\label{sec:rep}

\begin{figure}[b]
\includegraphics[width=1\columnwidth]{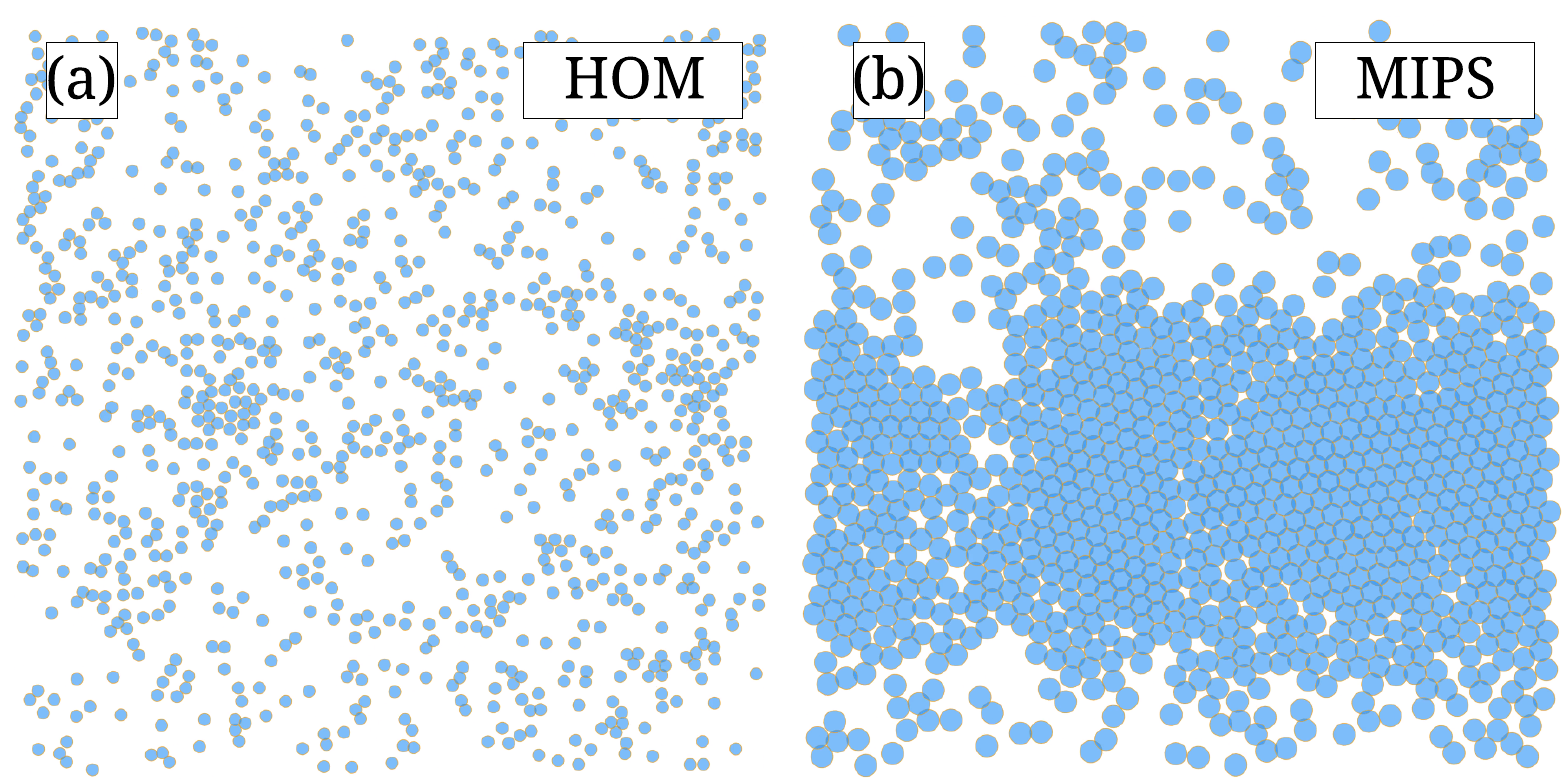}
\caption{\label{fig:MBcase} Simulation snapshots of the repulsive active Brownian particles showing two distinct phases determined by the particle size $\alpha$. (a)~Small size of particles ($\alpha v/D = 2 \times 10^3$) results in a homogeneous phase (HOM), whilst (b)~larger size of particles ($\alpha v/D = 4 \times 10^3$) leads to motility-induced phase separation (MIPS).}
\end{figure}

Having successfully tested our framework on a rather simple model, we now apply it to a many-body active system which features richer physics. Specifically, we consider a system of $N$ RABPs in two spatial dimensions that can exhibit MIPS [Fig~\ref{fig:MBcase}]\cite{Cates2015}. We take the dynamics in Eq.~\eqref{Con:eq:sec1:Langevin} with the potential energy given by
\begin{equation} \label{equ:MBpotentialenergy}
	\phi = \frac{1}{2} \ssum{i}{j}{N} U_{ij} ,
\end{equation}
where the pair potential $U_{ij}$ depends on the inter-particle distance $r_{ij} = |{\bf r}_i - {\bf r}_j|$. To impose repulsive interactions, we take $U_{ij} = \varepsilon \exp \big[ - 1/(1 - (r_{ij} / \alpha)^2) \big]$ for $r_{ij}<\alpha$, and $U_{ij}=0$ otherwise. The control parameter $\alpha$ here embodies the range of the repulsive interaction, which is akin to the excluded-volume radius. Therefore, changing $\alpha$ at a fixed number of particles amounts to changing the packing fraction, which can lead to a transition between homogeneous and phase-separated states [Fig~\ref{fig:MBcase}]. Moreover, we define the self-propulsion of particle $i$ as
\begin{equation} \label{equ:theta}
	\mathbf{v}_i = v_0 \,(\cos \theta_i, \sin\theta_i) ,
	\quad
	\dot{\theta}_i = \sqrt{2/\tau} \zeta_i ,
\end{equation}
where $v_0$ is the constant magnitude of self-propulsion speed, and $\zeta_i$ is a Gaussian white noise with zero mean and unit variance. The persistence time $\tau$, which controls the angular noise, determines the self-propulsion correlations as $\langle v_{in}(t) v_{im}(0)\rangle = \delta_{ij}\delta_{nm}v_0^2 e^{-|t|/\tau} $, where $n$ and $m$ refer to spatial coordinates~\cite{Fodor2018}.

Due to the difficulty in deriving analytical expressions for the decomposition of heat [Eqs.~(\ref{Con:eq:sec1:mainHeat}-\ref{equ:Greeksmain})], we resort to numerically evaluating the corresponding averages and correlations given in Appendix~\ref{Appendix:MBcase}. We perform simulations at different particle sizes $\lbrace \alpha \rbrace$ in the homogeneous case ($2.1\times 10^3<\alpha v/D<2.4\times 10^3$), and in the case of phase separation ($3.0\times 10^3<\alpha v/D<3.3\times 10^3$). After obtaining data for various $\alpha$ values, we fit averages and correlations to deduce their continuous dependence in terms of $\alpha$. We then use these fitting functions to obtain the control protocols through solving Eq.~\eqref{equ:tp}, where again we distinguish protocols for the homogeneous and MIPS cases. In practice, we consider $N = 2000$ for RABPs in a square box of side length $L = 200$, with periodic boundary conditions to mimic bulk conditions. The parameters are $v=1$, $\mu=1$, $D=10^{-3}$, $\tau=10^{2}$, and $\varepsilon=10^2 T$. Considering that lengthscales and timescales are respectively taken in units of micrometers and seconds, the parameter values are chosen to be relevant to actual microswimmers~\cite{Fodor2018}, such as moving bacteria. We integrate the equation of motions [Eqs.~\eqref{Con:eq:sec1:Langevin}~and~\eqref{equ:theta}] using a standard Euler discrete-update rule.

\begin{figure}
\includegraphics[width=1\columnwidth]{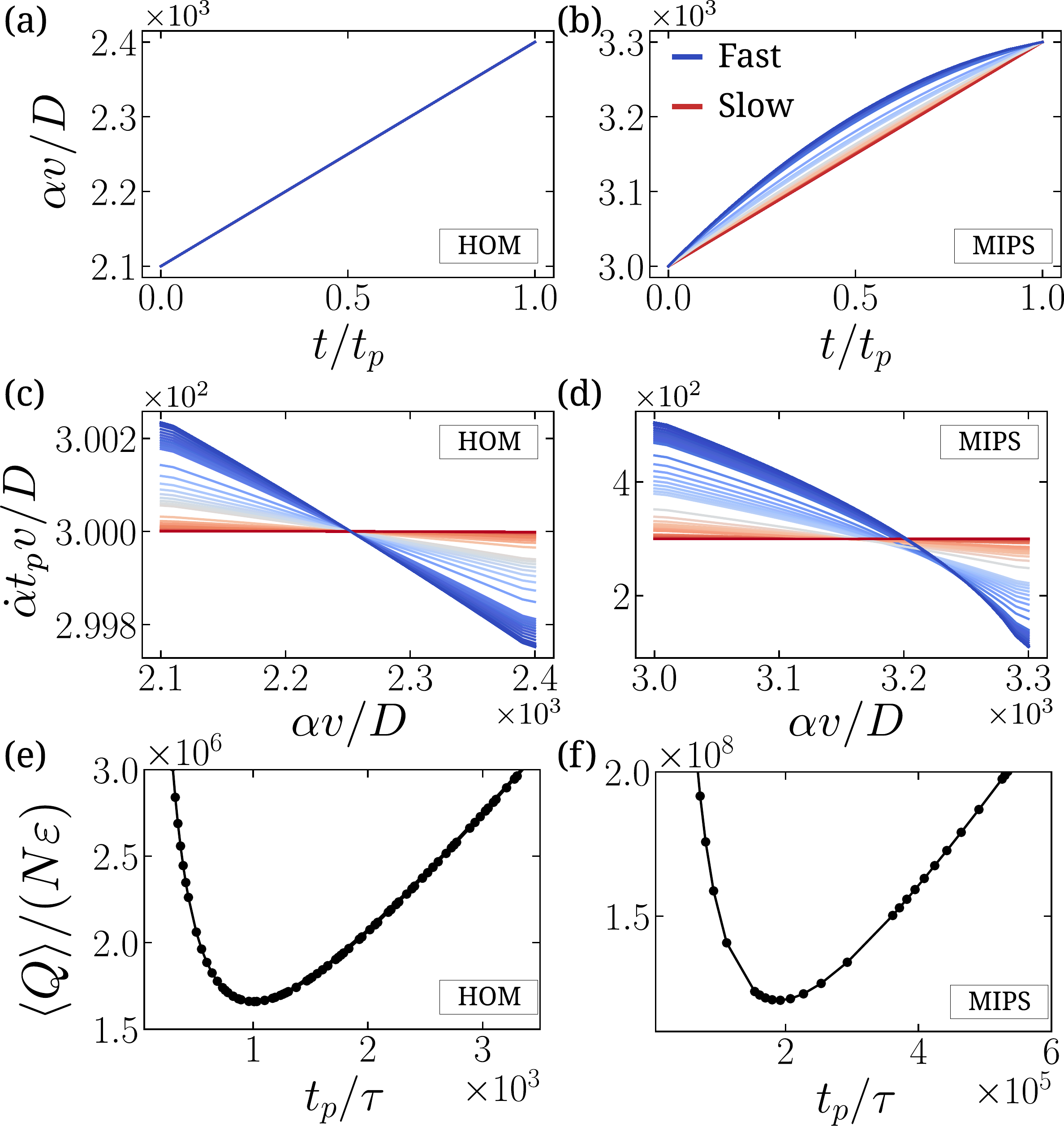}
\caption{\label{fig:MBcaseQ} Optimal thermodynamic control of a many-body system consisting of repulsive active Brownian particles in the homogeneous (HOM) (left) and phase-separated (MIPS) (right) cases. (a-b)~Derived protocols for driving the particle size $\alpha$ as a function of time. (c-d)~Control speed $\dot{\alpha}$ against control parameter $\alpha$. (e-f)~Scaled heat against protocol duration, computed using protocols shown in (a-b).}
\end{figure}

The resulting control protocols are shown in Figs.~\ref{fig:MBcaseQ}(a-b). In the homogeneous case, the control protocols \textit{appear} to lie on a single master curve, as for the control of passive systems~\cite{Deffner2020}. In contrast, for the MIPS case, we obtain various protocol curves for different protocol durations $t_p$, which collapse into distinct master curves either for very fast and very slow protocols. Interestingly, plotting the control speed $\dot\alpha$ against the value of the control parameter $\alpha$ reveals that the protocol curves do not actually collapse onto a single master curve in the homogeneous case [Fig.~\ref{fig:MBcaseQ}(c)], although the change in control speed is clearly smaller than that of the MIPS case [Fig.~\ref{fig:MBcaseQ}(d)]. Since the existence of a single master curve is a signature of passive systems~\cite{Deffner2020}, we speculate that RAPBs in the homogeneous state have \textit{macroscopic} signatures that resemble a thermal system, at least as far as the heat decomposition [Eqs.~(\ref{Con:eq:sec1:mainHeat}-\ref{equ:Greeksmain})] is concerned. Finally, we compute the heat as predicted from the response framework [Eq.~\eqref{Con:eq:sec1:mainHeat}]. The optimal protocol duration $t_p^\ast/\tau \approx 1 \times 10^3$ in the homogeneous case [Fig.~\ref{fig:MBcaseQ}(e)] is approximately 200 times larger than in the MIPS case, where $t_p^\ast / \tau \approx 2 \times 10^{5}$ [Fig.~\ref{fig:MBcaseQ}(f)]. This difference illustrates that the response to changes in particle size is slower in the MIPS case, where the cluster size adapts to the packing fraction, than in the homogeneous case.

% ===========================================================================================

\section{Discussion}

We have derived a thermodynamic framework for the optimal control of active systems that operate arbitrarily far from thermal equilibrium. The systematic nature of our approach relies on recent advances in stochastic thermodynamics and response theory, as inspired by previous works on the thermodynamic control of passive systems~\cite{Crooks2007, Sivak2012}. Applications of our framework have revealed new insights into the control of active systems that are in direct contrast to the passive case. Specifically, we have found that: (i) for non-zero activity, expanding the dissipation to linear order in the protocol duration requires knowledge of the second-order response, whereas linear response is sufficient to obtain the same order in the passive case~\cite{Sivak2012}; (ii) there is a cost (dissipation) associated with the self-driving meaning that the optimal duration is \textit{not} the longest one, at variance with passive systems~\cite{Sivak2012,Large2019}; (iii) we find that each protocol duration gives rise to a unique protocol curve which eventually collapse onto either a fast or slow master curve, in contrast to the strict many-to-one mapping as seen in passive control~\cite{Deffner2020, Zhong2022}.

As a demonstration of the generality of our approach, we have built our framework to describe both continuous and discrete-state active systems. This opens the door to future work that may quantitatively compare optimal control scenarios between these different categories of active matter. Interestingly, different definitions of irreversibility and of dissipation have been proposed for various models of active matter~\cite{Fodor2016, Speck2016, Gaspard2019, Gaspard2020, FodorCates2021}. Indeed, the measure of irreversibility provides a legitimate quantification of dissipation only if the underlying dynamics satisfies specific thermodynamic constraints. Our results on discrete-state dynamics can be regarded as a motivation to develop active discrete-state models that satisfy these constraints thermodynamically. Moreover, our approach could also be adapted to field theories of active matter~\cite{Nardini2017, Chate2020, Borthne2020}, some of which can be embedded within linear irreversible thermodynamics~\cite{Markovich2021}. Furthermore, although we have focused on optimizing heat, our framework can be straightforwardly adapted to other cost functions, for instance the extracted work (which has been optimized in the literature of passive systems~\cite{Schmiedl2007,GomezMarin2008,Sivak2012,Zulkowski2012,Zulkowski2013,Bonanca2014,Rotskoff2015,Rotskoff2017,Deffner2020,Zhong2022, Blaber2023}). Note that a recent study has considered optimizing the work in active matter by directly applying the framework of passive systems~\cite{Gupta23}.

Our framework relies on the assumption of weak and slow driving of the control parameter. It leads to an explicit decomposition of heat which can serve as a seed, or a testing ground, for machine learning approaches beyond the smooth-driving assumptions~\cite{Falk2021,Engel2022}. Interestingly, one could extend our framework to take into account corrections from higher-order responses, and thus to capture faster driving, within a systematic approach. Moreover, although we have focused here on controlling a single parameter for simplicity, it would be interesting to consider an arbitrary number of control variables in future works. Lastly, encouraged by the experimental implementations of the control framework in passive systems~\cite{Tafoya2019, Scandi2022} and recent progress in measuring the response of living systems~\cite{Toyabe2010, Turlier2016, Ahmed2018, Mizuno2018}, our framework can be readily deployed by experimentalists to explore the control of experimental active systems. Indeed, even without knowing the nonequilibrium response-correlation relation in Sec.~\ref{sec:responses}, our decomposition of heat in Eqs.~(\ref{Con:eq:sec1:mainHeat}-\ref{equ:Greeksmain}) already delineates which perturbation to apply and which response to measure. We anticipate that the potential success of applications of our framework will heavily depend on the quality of measured response functions.

Overall, our work paves the way towards designing active materials able to optimally switch between collective states. Indeed, while most studies of active matter focus on establishing phase diagrams, by associating control parameter values with collective states~\cite{Marchetti2013, Gompper2020}, how to optimally induce transitions has remained largely unexplored. While our framework sets the stage to this end, there remains a series of open challenges. First, the assumption of slow driving precludes crossing a critical transition, since the relaxation of the system is always slower than any perturbation at criticality~\cite{Campo2014}. Considering non-critical transitions instead, the corresponding order parameter is typically discontinuous, and so are the averages and correlations in the heat decomposition. Accordingly, one expects that the optimal protocol is no longer smooth when crossing phase boundaries, which challenges the assumption of weak driving. Therefore, it remains to be explored how the assumptions underlying the response framework need to be adapted when switching between collective states in active matter.

\acknowledgments{We thank Adolfo del Campo, Alessandro Manacorda, Tomer Markovich, and Elsen Tjhung for insightful discussions. L.K.D thanks the high performance computing team at the University of Luxembourg. K.P. is funded by the European Union’s Horizon 2020 research and innovation program under the Marie Sklodowska-Curie grant agreement No.~847523 ‘INTERACTIONS’, and from the Novo Nordisk Foundation (grant No. NNF18SA0035142 and NNF21OC0071284). L.K.D. and \'E.F. were supported through the Luxembourg National Research Fund (FNR), grant reference 14389168.}

% ===========================================================================================

\appendix

\section{Decomposition of heat}
\label{Appendix:heatinresponse}

In this Appendix, we derive the decomposition of heat given in Eqs.~(\ref{Con:eq:sec1:mainHeat}-\ref{equ:Greeksmain}). To this end, we assume that the control parameter $\alpha$ is varying both weakly and slowly, which allows us to rely on (i) the representation of averages in terms of response functions [Eq.~\eqref{eq:resp}], and (ii) the expansion of the control parameter variation in terms of time increments [Eq.~\eqref{eq:increm}]. Using these approximations, we first simplify the averages featuring in the heat definition [Eq.~\eqref{equ:heatraw}]. We first expand $\langle \phi_1 \rangle$ as
\begin{equation}\label{eq:avg_1}
	\begin{aligned}
		\langle \phi_1 \rangle &\approx \langle \phi_1 \srangle + \int_{-\infty}^{t_p} dt' \Delta \alpha_{t_p,t'} R_1(\phi; t_p, t'),
		\\
		&\approx \langle \phi_1 \srangle + \dot\alpha_1 \int_{0}^{\infty} dt R_1(\phi; t, 0) t ,
	\end{aligned}
\end{equation}
where we have used the approximations (i) and (ii) in the first and second lines, respectively. Similarly, we deduce $\langle \partial_\alpha\phi \rangle$ as
\begin{equation}\label{eq:avg_2}
	\begin{aligned}
		\langle \partial_\alpha\phi(t) \rangle &\approx \langle \partial_\alpha\phi\srangle + \int_{-\infty}^{t} dt' \Delta \alpha_{t,t'} R_1(\partial_\alpha\phi; t, t')
		\\
		&\approx \langle \partial_\alpha\phi \srangle + \dot\alpha(t) \int_{0}^{\infty} dt' R_1(\partial_\alpha\phi; t', 0) t' .
	\end{aligned}
\end{equation}
Finally, $\langle J\rangle$ follows as
\begin{equation}\label{eq:avg_3}
	\begin{aligned}
		\langle J(t) \rangle &\approx \langle J\srangle + \int_{-\infty}^{t} dt' \Delta \alpha_{t,t'} R_1(J; t, t')
		\\
		&\quad + \int_{-\infty}^{t} \int_{-\infty}^{t} dtdt' \Delta \alpha_{t,t'} \Delta \alpha_{t,t''} R_2(J; t,t',t''),
		\\
		&\approx \langle J \srangle + \dot\alpha(t) \int_{0}^{\infty} dt' R_1(J; t', 0) t'
		\\
		&\quad + \frac{\ddot\alpha(t)}{2} \int_{0}^{\infty} dt' R_1(J; t', 0) {t'}^2
		\\
		&\quad + \dot\alpha(t)^2 \int_{0}^{\infty} \int_{0}^{\infty} dt' dt'' R_2(J; t',t'-t'',0) t' t'' .
	\end{aligned}
\end{equation}
In contrast with Eqs.~(\ref{eq:avg_1}-\ref{eq:avg_2}), for Eq.~\eqref{eq:avg_3} we have used the second-order response $R_2$, and also the expansion of $\Delta \alpha$ up to second order in $\Delta t$ [Eq.~\eqref{eq:increm}]. These higher-order corrections are needed for consistency in the overall expansion of the heat, to leading order in $\dot\alpha$. Substituting Eqs.~(\ref{eq:avg_1}-\ref{eq:avg_3}) into Eq.~\eqref{equ:heatraw}, then integrating the second last term in Eq.~\eqref{eq:avg_3} by parts, we deduce the expression of the boundary and Lagrangian terms in the heat [Eqs.~(\ref{Con:eq:sec1:mainHeat}-\ref{equ:Greeksmain})].

% ------------------------------------------------------------------------------------------

\section{Scalings of heat}
\label{Appendix:scalingtp}

In this Appendix, we examine how the heat $\langle Q\rangle$ scales with the protocol time $t_p$ by considering the asymptotic regimes for fast protocols ($t_p \ll t_p^\ast$) and slow protocols ($t_p \gg t_p^\ast$), where $t_p^*$ minimizes $\langle Q\rangle$ [Fig.~\ref{fig:Sketch}]. We rely on the decomposition $\langle Q\rangle = B + \int dt{\cal L}$ given in terms of the boundary and Lagrangian contributions [Eqs.~(\ref{Con:eq:sec1:mainHeat}-\ref{equ:Greeksmain})], respectively denoted by $B$ and $\cal L$. Besides, we express $\cal L$ as
\begin{equation}\label{eq:B1}
	{\cal L} = E - 2V ,
\end{equation}
where we have introduced the constant of motion $E$, defined in Eq.~\eqref{equ:1stIntegral}.

First, it appears from Eqs.~(\ref{equ:controlspeed}-\ref{equ:tp}) that the protocol duration $t_p$ is small whenever the constant of motion $E$ satisfies $E\gg\underset{\alpha}{\max} \,V$. For fast protocols ($t_p\ll t_p^*$), we can then simplify Eqs.~(\ref{equ:controlspeed}-\ref{equ:tp}) as
\begin{equation}\label{eq:E_sc}
	\begin{aligned}
		E &\underset{t_p\ll t_p^*}{\sim} \frac{1}{t_p^2} \bigg[\int_{\alpha_0}^{\alpha_1} d\alpha \sqrt{ (\Psi-\Sigma')(\alpha) } \bigg]^2 ,
		\\
		\dot\alpha &\underset{t_p\ll t_p^*}{\sim} \pm \sqrt{ \frac{E}{(\Psi-\Sigma')(\alpha)} } .
	\end{aligned}
\end{equation}
Substituting the expression of $E$ into that of $\dot\alpha$, we deduce after separating variables and integration that the optimal protocol obeys
\begin{equation}
	\int_{\alpha_0}^{\alpha(t)} d\alpha \sqrt{ (\Psi-\Sigma')(\alpha) } \underset{t_p\ll t_p^*}{\approx} \frac{t}{t_p} \int_{\alpha_0}^{\alpha_1} d\alpha \sqrt{ (\Psi-\Sigma')(\alpha) } ,
\end{equation}
where we have assumed $\alpha_0<\alpha_1$. Therefore, it follows that the optimal trajectory $\alpha(t)$ follows a master curve when scaling $t$ with $t_p$, as for the control of passive systems~\cite{Sivak2012}. Moreover, the condition $E\gg\underset{\alpha}{\max} \,V$ yields
\begin{equation} \label{eq:B2}
	\int_0^{t_p} dt {\cal L} \underset{t_p\ll t_p^\ast}{\sim} E \,t_p ,
\end{equation}
and we express the boundary term [Eq.~\eqref{eq:bound}], using the simplified expression of $\dot\alpha$ [Eq.~\eqref{eq:E_sc}], as
\begin{equation} \label{eq:B4}
	B \underset{t_p\ll t_p^\ast}{\sim} \sqrt{E} \,\bigg[ \frac{\Sigma}{\sqrt{\Psi-\Sigma'}}(\alpha_0) - \frac{\Sigma+\Phi}{\sqrt{\Psi-\Sigma'}} (\alpha_1) \bigg] ,
\end{equation}
where we have ignored terms that do not depend on $t_p$. Finally, combining Eqs.~(\ref{eq:B2}-\ref{eq:B4}) with the fact that $E\sim 1/t_p^2$ [Eq.~\eqref{eq:E_sc}] results in the first scaling relation for the heat:
\begin{equation}
\langle Q \rangle \underset{t_p\ll t_p^\ast}{\sim} K(\alpha_0, \alpha_1)/t_p ,
\end{equation}
where
\begin{equation} \label{equ:Ka0a1}
	\begin{aligned}
		K(\alpha_0,\alpha_1) &= c(\alpha_0,\alpha_1) \bigg[\frac{\Sigma}{\sqrt{\Psi-\Sigma'}}(\alpha_0) -\frac{\Sigma+\Phi}{\sqrt{\Psi-\Sigma'}}(\alpha_1) \bigg]
		\\
		&\quad + c(\alpha_0,\alpha_1)^2 ,
		\\
		c(\alpha_0,\alpha_1) &= \int_{\alpha_0}^{\alpha_1} d\alpha \sqrt{(\Psi-\Sigma')(\alpha)} .
	\end{aligned}
\end{equation}
The term $K$ accounts for contributions from both the boundary and Lagrangian terms.

Second, we deduce from Eqs.~(\ref{equ:controlspeed}-\ref{equ:tp}) that the protocol duration $t_p$ is large whenever $E\approx\underset{\alpha}{\max} \,V$. Therefore, for slow protocols ($t_p\gg t_p^*$), the constant of motion $E$ depends only weakly on the protocol duration $t_p$. It follows from Eq.~\eqref{equ:controlspeed} that the protocol speed $\dot\alpha(t)$ depends weakly on $t_p$, and so does the boundary term $B$ [Eq.~\eqref{eq:bound}]. Plugging the condition $E\approx\underset{\alpha}{\max} \,V$ into Eq.~\eqref{eq:B1}, and integrating over the protocol results in 
\begin{equation}
	\int_0^{t_p} dt {\cal L} \underset{t_p\gg t_p^\ast}{\sim} - t_p \max_\alpha V.
\end{equation}
Therefore, we arrive at the second scaling relation for the heat:
\begin{equation}
    \langle Q\rangle \underset{t_p \gg t_p^\ast}{\sim} t_p \min_\alpha \langle J\srangle ,
\end{equation}
where we have used $V = - \langle J \srangle$.

% ------------------------------------------------------------------------------------------

\section{Response theory from path probability}

In this Appendix, we show how to explicitly relate response and correlation functions using a path probability representation of the dynamics. We consider two cases: (i) continuous-state dynamics [Eq.~\eqref{Con:eq:sec1:Langevin}], and (ii) discrete-state dynamics [Eq.~\eqref{equ:Compactmaster}].

\subsection{Continuous-state dynamics}\label{app:ccos}

The linear response function $R_1$, defined in Eq.~\eqref{eq:resp_b}, can be written in terms of the dynamic action $\cal S$ [Eq.~\eqref{eq:action}] as
\begin{equation}\label{equ:Appendix:explicitR1}
	\begin{aligned}
		&R_1({X}; t, t')
		\\
		&= \int  {X}(t)  \bigg[ \frac{\delta \mathcal{N} }{\delta \Delta \alpha_{t,t'}} - {\cal N}\frac{\delta \mathcal{S}}{\delta \Delta \alpha_{t,t'}}\bigg]_{\Delta\alpha\to 0} e^{-{\cal S}}\,\mathcal{D} [{\bm\eta}, {\bf v}] ,
	\end{aligned}
\end{equation}
where ${\cal N} = 1/\int e^{-{\cal S}} \mathcal{D} [{\bm\eta}, {\bf v}]$. Given that
\begin{equation}
	\begin{aligned}
		 \frac{\delta \mathcal{N}}{\delta \Delta \alpha}\bigg|_{\Delta\alpha\to 0} &= {\cal N}^2 \int \frac{\delta \mathcal{S}}{\delta \Delta \alpha}\bigg|_{\Delta\alpha\to 0} e^{-{\cal S}}\,\mathcal{D} [{\bm\eta}, {\bf v}]
		\\
		&= {\cal N} \left\langle \frac{\delta\mathcal{S}}{\delta \Delta \alpha} \right\srangle \bigg|_{\Delta\alpha\to 0} ,
	\end{aligned}
\end{equation}
we deduce
\begin{equation}
\begin{aligned}\label{eq:R1_con}
       R_1({X}; t, t') &= \bigg[ \langle X \srangle \left\langle \frac{\delta\mathcal{S}}{\delta \Delta \alpha} \right\srangle  - \left\langle X(t) \frac{\delta\mathcal{S}}{\delta \Delta \alpha_{t,t'}}  \right\rangle \bigg]_{\Delta\alpha\to 0}  \\
       & = -\llangle[\bigg] X(t)  \frac{\delta \mathcal{S} }{\delta \Delta \alpha_{t,t'}} \rrangle[\bigg]_{\Delta\alpha\to 0},
\end{aligned}
\end{equation}
where we have used Eq.~\eqref{eq:2pt}. From Eq.~\eqref{eq:action}, we get
\begin{equation} \label{equ:Appendix:Sd1}
	\begin{aligned}
		\frac{\delta \mathcal{S}}{\delta \Delta \alpha}\bigg|_{\Delta\alpha\to 0} &= - \frac{1}{2T} \bigg[ (\nabla_i\partial_\alpha\phi) \cdot ( \dot{\bf r}_i - \mu{\bf f}_i ) \bigg] 
		\\
		&= - \frac{1}{2T} \bigg[ \partial_\alpha\dot\phi - \mu (\nabla_i\partial_\alpha\phi) \cdot {\bf f}_i ) \bigg] ,
	\end{aligned}
\end{equation}
where we have used the chain rule $\partial_\alpha \dot \phi = \dot{\bf r}_i\cdot\nabla_i \partial_\alpha \phi$ in the unperturbed dynamics (namely, at constant $\alpha$). Finally, substituting Eq.~\eqref{equ:Appendix:Sd1} into Eq.~\eqref{eq:R1_con} yields the expression for the linear response function
\begin{equation}\label{equ:R1cont_app}
	2T R_1(X;t,t') = \frac{d}{dt'} \llangle[\big]{X}_t [\partial_\alpha \phi]_{t'} \rrangle[\big] - \mu \llangle[\big] X_t \big[ (\nabla_i \partial_\alpha \phi)\cdot \mathbf{f}_i \big]_{t'} \rrangle[\big] ,
\end{equation}
in agreement with in Eq.~\eqref{equ:R1cont}. Symmetrizing the linear response function results in the following
\begin{equation}
  \begin{aligned}
      &R_1(X;t,t') - R_1(X;t',t)
      \\
      &\quad=\frac{1}{2T}\bigg[\frac{d}{dt'} \llangle X_t [\partial_\alpha \phi]_{t'} \rrangle - \mu \llangle X_t [(\nabla_i \partial_\alpha \phi) \cdot \mathbf{f}_i]_{t'} \rrangle \\
      &\qquad -\frac{d}{dt} \llangle X_{t'} [\partial_\alpha \phi]_{t} \rrangle + \mu \llangle X_{t'} [(\nabla_i \partial_\alpha \phi) \cdot \mathbf{f}_i]_{t} \rrangle \bigg] .
  \end{aligned}
\end{equation}
In the passive case ($\mathbf{v} = 0$), the force acting on particles reduces to ${\bf f}_i = -\nabla_i\phi$. Besides, time-reversal symmetry yields $\llangle X_t Y_{t'} \rrangle  = \llangle X_{t'} Y_t \rrangle$ and $(d/dt) \llangle X_{t'} Y_{t} \rrangle = -(d/dt') \llangle X_t Y_{t'} \rrangle $ for time-symmetric observables $X$ and $Y$~\cite{Onsager1931}. Causality enforce that $R_1(X;t',t) = 0$ for $t' < t$, from which we deduce
\begin{equation} 
	R(X;t,t')=\frac{1}{T}\frac{d}{dt'} \llangle X_t [\partial_\alpha \phi]_{t'} \rrangle ,
\end{equation}
as predicted by the fluctuation-dissipation theorem~\cite{Kubo1966}.

Similarly, we can write the second-order response function $R_2$, defined in Eq.~\eqref{eq:resp_b}, as
\begin{equation}
	\begin{aligned}
		R_2&({X}; t, t', t'')
		\\
		&= \int  {X}(t) \bigg[ {\cal N} \bigg( \frac{\delta \mathcal{S}}{\delta \Delta \alpha_{t,t'}} \frac{\delta \mathcal{S}}{\delta \Delta \alpha_{t,t''}} - \frac{\delta^2 \mathcal{S}}{\delta \Delta \alpha_{t,t'}\delta \Delta \alpha_{t,t''}} \bigg)
		\\
		&\quad - \frac{\delta \mathcal{N}}{\delta \Delta \alpha_{t,t''}} \frac{\delta \mathcal{S}}{\delta \Delta \alpha_{t,t'}} - \frac{\delta \mathcal{N}}{\delta \Delta \alpha_{t,t'}} \frac{\delta \mathcal{S}}{\delta \Delta \alpha_{t,t''}}
		\\
		&\quad + \frac{\delta^2 \mathcal{N}}{\delta \Delta \alpha_{t,t'}\delta \Delta \alpha_{t,t''}} \bigg]_{\Delta\alpha\to 0} e^{-{\cal S}}\,\mathcal{D} [{\bm\eta}, {\bf v}] .
	\end{aligned}
\end{equation}
Given that
\begin{equation}
	\begin{aligned}
		&\frac{\delta^2 \mathcal{N}}{\delta \Delta \alpha_{t,t'} \delta \Delta \alpha_{t,t''}}\bigg|_{\Delta\alpha\to 0}
		\\
		&\quad= {\cal N} \,\bigg[ \left\langle \frac{\delta^2\mathcal{S}}{\delta \Delta \alpha_{t,t'}\delta \Delta \alpha_{t,t''}} - \frac{\delta\mathcal{S}}{\delta \Delta \alpha_{t,t'}} \frac{\delta\mathcal{S}}{\delta \Delta \alpha_{t,t''}} \right\rangle
 \\
		&\qquad + 2 \left\langle \frac{\delta\mathcal{S}}{\delta \Delta \alpha_{t,t'}} \right\srangle \left\langle \frac{\delta\mathcal{S}}{\delta \Delta \alpha_{t,t''}} \right\srangle \bigg]_{\Delta\alpha\to 0} ,
	\end{aligned}
\end{equation}
we deduce
\begin{equation} \label{equ:Appendix:explicitR2}
	\begin{aligned}
		R_2({X};t,t',t'')	&= \lllangle[\bigg] X(t) \bigg[ \frac{\delta \mathcal{S}}{\delta \Delta \alpha_{t,t'}} \frac{\delta \mathcal{S}}{\delta \Delta \alpha_{t,t''}}
		\\
		&\quad\qquad - \frac{\delta^2 \mathcal{S}}{\delta \Delta \alpha_{t,t'}\delta \Delta \alpha_{t,t''}} \bigg] \rrrangle[\bigg]_{\Delta\alpha\to 0},
	\end{aligned}
\end{equation}
where we have used Eq.~\eqref{eq:3pt}. From Eq.~\eqref{eq:action}, we get
\begin{equation} \label{equ:Appendix:Sd2}
	\begin{aligned}
		&\frac{\delta^2 \mathcal{S}}{\delta \Delta \alpha_{t,t'}\delta \Delta \alpha_{t,t''}} \bigg|_{\Delta\alpha\to 0}
		\\
		&\; = \frac{\delta(t'-t'')}{2T} \Big[ \mu (\nabla_i \partial_\alpha \phi)^2 + (\nabla_i \partial_\alpha^2 \phi)\cdot ( \dot{\mathbf{r}}_i - \mu \mathbf{f}_i) \Big] .
	\end{aligned}
\end{equation}
Finally, substituting Eqs.~\eqref{equ:Appendix:Sd1} and~\eqref{equ:Appendix:Sd2} into Eq.~\eqref{equ:Appendix:explicitR2} yields the expression for the second-order response function given in Eq.~\eqref{equ:R2cont}.

% ------------------------------------------------------------------------------------------

\subsection{Discrete-state dynamics}\label{app:diss}

We can repeat the above derivation for the case of discrete-state dynamics. The response functions can be written in terms of perturbations of the dynamic action ${\cal A}$ [Eq.~\eqref{eq:dadis}] as
\begin{equation}\label{eq:R1discrete}
	\begin{aligned}
    R_1({X}; t, t') &= -\llangle[\bigg] X(t)  \frac{\delta \mathcal{A}}{\delta \Delta \alpha_{t,t'}} \rrangle[\bigg]_{\Delta\alpha\to 0} ,
    \\
		R_2({X};t,t',t'')	&= \lllangle[\bigg] X(t) \bigg[ \frac{\delta \mathcal{A}}{\delta \Delta \alpha_{t,t'}} \frac{\delta \mathcal{A}}{\delta \Delta \alpha_{t,t''}}
		\\
		&\quad\qquad - \frac{\delta^2 \mathcal{A}}{\delta \Delta \alpha_{t,t'}\delta \Delta \alpha_{t,t''}} \bigg] \rrrangle[\bigg]_{\Delta\alpha\to 0} .
	\end{aligned}
\end{equation}
From Eq.~\eqref{eq:dadis}, we get
\begin{equation}\label{eq:da1discrete}
	\begin{aligned}
		\frac{\delta \mathcal{A}}{\delta \Delta \alpha}\bigg\vert_{\Delta \alpha\rightarrow 0} &= \frac{1}{2T} \sum_{i,j \neq i} \Big[ K_{ij}\rho_j \partial_\alpha \phi_i - n_{ij} \partial_\alpha(\phi_i-\phi_j) \Big]
		\\
		&= \frac{1}{2T} \bigg( \sum_{i,j \neq i} K_{ij}\rho_j \partial_\alpha \phi_i - \partial_\alpha \dot \phi \bigg) , 
	\end{aligned}
\end{equation}
where we have used that $\sum_{i}K_{ij}=0$, which follows from $\sum_i \dot p_i=0$ and Eq.~\eqref{equ:Compactmaster}, and also $\partial_\alpha\dot\phi = \sum_{i,j\neq i} n_{ij} \partial_\alpha(\phi_i-\phi_j)$ in the unperturbed dynamics. Similarly, one can use Eq.~\eqref{eq:dadis} to show that
\begin{equation}\label{eq:da2discrete}
	\begin{aligned}
		&\frac{\delta^2 \mathcal{A}}{\delta \Delta \alpha_{t,t'}\delta \Delta \alpha_{t,t''}} \bigg|_{\Delta\alpha\to 0}
		\\
		&= \frac{\delta(t'-t'')}{2T} \sum_{i,j } \Big[ K_{ij}\rho_j \Big( \partial^2_{\alpha}\phi_i - \frac{(\partial_{\alpha}\phi_j - \partial_{\alpha}\phi_i)^2}{2T} \Big)
		\\
		&\qquad + n_{ij} \partial^2_{\alpha}(\phi_i-\phi_j) \Big] .
	\end{aligned}
\end{equation}
Combining Eqs.~(\ref{eq:R1discrete}-\ref{eq:da2discrete}) gives the discrete-state response functions in Eqs.~(\ref{equ:R1disc}-\ref{equ:R2disc}).

% ------------------------------------------------------------------------------------------

\section{Applications of thermodynamic control}

In this Appendix, we determine how the generic decomposition of heat, given in Eqs.~(\ref{Con:eq:sec1:mainHeat}-\ref{equ:Greeksmain}) for arbitrary potential $\phi$ and self-propulsion ${\bf v}_i$, translates in two specific cases. We consider the two examples discussed in the main text: an active particle in harmonic trap [Sec.~\ref{sec:harm}], and an assembly of active particles with purely repulsive interactions [Sec.~\ref{sec:rep}].

%\begin{widetext}
\onecolumngrid
\subsection{Control of trap stiffness}
\label{Appendix:1Dcase}

In the case of an active particle in a one-dimensional harmonic trap, we have $\phi=\frac{1}{2} \alpha r^2$ for the potential energy, yielding: $\partial_\alpha \phi = \frac{1}{2}r^2$, $\nabla \phi = \alpha r$, and $\nabla \partial_\alpha \phi = r$. The functions in the decomposition of heat [Eqs.~(\ref{Con:eq:sec1:mainHeat}-\ref{equ:Greeksmain})] then read
\begin{equation}\label{equ:1Dallcorrs}
	\begin{aligned}
		\Sigma &= \frac{\alpha}{4T} \int_0^\infty dt \Bigg[ \llangle {}[rv]_t [rv]_0 \rrangle t^2 - t\left(\frac{1}{2} + \alpha \mu t\right) \llangle {}[rv]_t r^2_0 \rrangle  \Bigg],
		\\
		\Lambda &= \frac{1}{2} \langle r^2\srangle + \frac{ \alpha}{2T} \int_0^\infty dt \Bigg[ \llangle {}[rv]_t {}[rv]_0 \rrangle t- \left(\frac{1}{2} + \alpha \mu t\right)\llangle {}[rv]_t r^2_0 \rrangle \Bigg] ,
		\\
		\Phi &=  \frac{\alpha}{4T} \int_0^\infty dt \Bigg[ \left(\frac{1}{2} + \alpha \mu t\right) \llangle r^2_t r^2_0 \rrangle  - \llangle r^2_t [rv]_0 \rrangle t \Bigg] ,
	\end{aligned}	
\end{equation}
and for $\Psi$ we have
\begin{equation} \label{equ:1Dallcorrspsi}
	\begin{aligned}
		\Psi &= \frac{1}{4T} \int_0^\infty dt \bigg[ \left(\frac{1}{2} + \alpha \mu t \right) \llangle r^2_t r^2_0  \rrangle - \llangle r^2_t [rv]_0 \rrangle t - \alpha\mu t^2 \llangle {}[rv]_t r^2_{0}\rrangle \bigg]
		\\
		&\quad + \frac{\alpha }{8T^2} \int_0^\infty \int_0^\infty  dt dt'\bigg[ t'\left( \frac{1}{2}+ \mu \alpha t \right) \lllangle {}[rv]_t{}[rv]_{t-t'} r^2_0 \rrrangle + t \left( \frac{1}{2} + \mu \alpha t' \right) \lllangle {}[rv]_t r^2_{t-t'} [rv]_0 \rrrangle 
		\\
		&\quad - \left( \frac{1}{4} + (\alpha\mu)^2 tt' + \frac{\alpha \mu (t+t')}{2} \right)\lllangle {}[rv]_t r^2_{t-t'} r^2_0 \rrrangle - \alpha tt' \lllangle {}[rv]_t[rv]_{t-t'} [rv]_0 \rrrangle \bigg].
	\end{aligned}
\end{equation}
To obtain expressions for the correlation functions, we first solve the dynamics in Eq.~\eqref{eq:1dmodel} at constant $\alpha$, resulting in
\begin{equation}
	r(t) = \int_{-\infty}^{t} dt' e^{-\alpha \mu (t-t') } \Big[ \sqrt{2 D} \eta(t') + v (t') \Big] .
\end{equation}
From this solution, we can write any two-point (unconnected) correlation function. For instance
\onecolumngrid
   \begin{equation}\label{eq:1d:corr1}
	\begin{aligned}
		\langle {}[rv]_t r^2_{t'} \rangle &= \int_{-\infty}^{t} d\tau_1 \int_{-\infty}^{t'} d\tau_2 \int_{-\infty}^{t'} d\tau_3 e^{\alpha \mu (\sum_i \tau_i - t - 2t' )} \langle (\sqrt{2D} \eta_1 + v_1) v(t) (\sqrt{2D} \eta_2 + v_2) (\sqrt{2D} \eta_3 + v_3) \rangle
		\\
		&= \int_{-\infty}^{t} d\tau_1 \int_{-\infty}^{t'} d\tau_2 \int_{-\infty}^{t'} d\tau_3 e^{\alpha \mu (\sum_i \tau_i - t - 2t' )} \Big[ \langle  v_1 v(t) \rangle\langle v_2 v_3 \rangle + \langle v_1 v_2  \rangle \langle v(t) v_3 \rangle + \langle v_1 v_3 \rangle \langle v(t) v_2 \rangle
		\\
		&\quad + 2D \big( \langle  v_1 v(t) \rangle\langle \eta_2 \eta_3 \rangle + \langle \eta_1 \eta_2  \rangle \langle v(t) v_3 \rangle  + \langle \eta_1 \eta_3 \rangle \langle v(t) v_2 \rangle \big) \Big] ,
	\end{aligned}
\end{equation} 
where we have introduced the notations $\eta_i=\eta(\tau_i)$ and $v_i=v(\tau_i)$. We have used that $\eta$ and $v$ are independent Gaussian noises with zero mean. Finally, using that $\langle\eta(t)\eta(t')\rangle = \delta(t-t')$ and $\langle v(t)v(t')\rangle = (D_1/\tau) e^{-|t-t'|/\tau}$, one can compute all the integrals in Eq.~\eqref{eq:1d:corr1}. This approach carries over to all the two- and three-point correlation functions in Eq.~\eqref{equ:1Dallcorrs}. It leads to exact expressions for the decomposition of heat [Eqs.~(\ref{Con:eq:sec1:mainHeat}-\ref{equ:Greeksmain})], given in Eq.~\eqref{equ:1DGreeks} for the case of an active particle in harmonic trap. 
% ------------------------------------------------------------------------------------------

\subsection{Control of particle size}
\label{Appendix:MBcase}

For many-body dynamics with pairwise interactions, the potential takes the form $\phi = \sum_{i,j<i} U_{ij}$. To compute the functions in the decomposition of heat [Eqs.~(\ref{Con:eq:sec1:mainHeat}-\ref{equ:Greeksmain})], the following quantities are needed:
\begin{equation}\label{equ:As}
	\begin{aligned}
		A_1  &= \frac{1}{2} \ssum{i}{j}{N} U_{ij} , 
		\quad
		A_2 = \frac{1}{2} \ssum{i}{j}{N} \partial_\alpha U_{ij} ,
		\quad
		A_3 =\ssum{i}{j}{N} \mathbf{v}_i \cdot \partial_{{\bf r}_i} U_{ij} ,
		\quad
		A_4 = \sum _{i,j\neq i} (\partial_{{\bf r}_i} \partial_\alpha U_{ij}) \cdot \Big( \mu \partial_{{\bf r}_i} \sum_{k\neq i} U_{ik} - \mathbf{v}_i \Big) ,
		\\
		A_5 &= \frac{1}{2} \ssum{i}{j}{N} \partial_{\alpha}^2 U_{ij} ,
		\quad
		A_6 = \sum_{i,j\neq i} ( \partial_{{\bf r}_i} \partial_{\alpha}^2 U_{ij} ) \cdot \Big( \mu \partial_{{\bf r}_i} \sum_{k\neq i} U_{ik} - \mathbf{v}_i \Big) ,
		\quad
		A_7 = \sum _{i,j\neq i,k\neq i} ( \partial_{{\bf r}_i} \partial_{\alpha} U_{ij} ) \cdot ( \partial_{{\bf r}_i} \partial_{\alpha} U_{ik} ) ,
	\end{aligned}
\end{equation}
from which we deduce
\begin{equation} \label{equ:MBgreeks1}
	\begin{aligned}
		V &= \langle A_3 \srangle ,
		\\
        \Phi &= \frac{1}{2T} \int_0^\infty dt \Big[ \llangle A_1(t) A_2(0) \rrangle + t \llangle A_1(t) A_4(0) \rrangle \Big],
        \\
		\Sigma &= \frac{1}{4T} \int_0^\infty  dt \Big[ t\llangle A_3(t) A_2(0) \rrangle+ t^2 \llangle A_3(t) A_4(0) \rrangle \Big] ,
		\\
		\Lambda &= \langle A_2 \srangle - \frac{1}{2T} \int_0^\infty dt \Big[  \llangle A_3(t) A_2(0) \rrangle + t\llangle A_3(t) A_4(0) \rrangle \Big],
    \end{aligned}
\end{equation}
and $\Psi$ can be expressed as
\begin{equation} \label{equ:MBgreeks2}
	\begin{aligned}
		\Psi &= \frac{1}{2T} \int_0^\infty dt \Big[ \llangle A_2(t) A_2(0) \rrangle + t\llangle A_2(t) A_4(0) \rrangle  - \frac{t}{2} \Big( \llangle A_3(t) A_5(0) \rrangle + t\llangle A_3(t) A_6(0) \rrangle + \mu t \llangle A_3(t) A_7(0) \rrangle \Big) \Big]
		\\
		&\quad - \frac{1}{8T^2} \int_0^\infty \int_0^\infty dtdt' \Big[\lllangle A_3(t) A_2(t-t') A_2(0) \rrrangle + t \lllangle A_3(t) A_2(t-t') A_4(0) \rrrangle + t' \lllangle A_3(t) A_4(t-t') A_2(0) \rrrangle
		\\
		&\quad + t t' \lllangle A_3(t) A_4(t-t') A_4(0) \rrrangle \Big].
    \end{aligned}
\end{equation}
%\end{widetext}

% ===========================================================================================

\twocolumngrid
%\bibliographystyle{physrev}
%\bibliography{note-ref-EF}
%merlin.mbs apsrev4-1.bst 2010-07-25 4.21a (PWD, AO, DPC) hacked
%Control: key (0)
%Control: author (0) dotless jnrlst
%Control: editor formatted (1) identically to author
%Control: production of article title (0) allowed
%Control: page (1) range
%Control: year (0) verbatim
%Control: production of eprint (0) enabled

%merlin.mbs apsrev4-1.bst 2010-07-25 4.21a (PWD, AO, DPC) hacked
%Control: key (0)
%Control: author (0) dotless jnrlst
%Control: editor formatted (1) identically to author
%Control: production of article title (0) allowed
%Control: page (1) range
%Control: year (0) verbatim
%Control: production of eprint (0) enabled
%

\end{document}